\documentclass[twocolumn,prl, twocolumn,superscriptaddress, nofootinbib]{revtex4-1}

\usepackage[T1]{fontenc}
\usepackage[latin9]{inputenc}
\usepackage{color}
\usepackage{textcomp}
\usepackage{mathrsfs}
\usepackage{amsmath}
\usepackage{amssymb}
\usepackage{graphicx}
\usepackage[unicode=true,pdfusetitle,
 bookmarks=true,bookmarksnumbered=false,bookmarksopen=false,
 breaklinks=false,pdfborder={0 0 0},pdfborderstyle={},backref=false,colorlinks=true]
 {hyperref}
\hypersetup{
 citecolor=blue,urlcolor=blue}

\makeatletter
\usepackage{bbm}
\usepackage[final]{changes}
\usepackage{ulem}

\makeatother

\begin{document}
\title{Do Majorana zero modes emerge in the hybrid nanowire under a strong magnetic field?}
\author{Guo-Jian Qiao }
\affiliation{Beijing Computational Science Research Center, Beijing 100193, China}
\author{Sheng-Wen Li}
\email{lishengwen@bit.edu.cn}

\affiliation{Center for Quantum Technology Research, and Key Laboratory of Advanced
Optoelectronic Quantum Architecture and Measurements, School of Physics,
Beijing Institute of Technology, Beijing 100081, People\textquoteright s
Republic of China}
\author{C. P. Sun}
\email{suncp@gscaep.ac.cn}

\affiliation{Beijing Computational Science Research Center, Beijing 100193, China}
\affiliation{Graduate School of China Academy of Engineering Physics, Beijing 100193,
China}
\begin{abstract}
The hybrid nanowire consisting of semiconductor with proximity to
superconductor is expected to serve as an experimental platform to
display Majorana zero modes. By rederiving its effective Kitaev model
with spins, we discover a novel topological phase diagram, which assigns
a more precise constraint on the magnetic field strength for the emergence
of Majorana zero modes. It then turns out the effective pairing strength
dressed by the proximity effect exhibits a significant dependence on the magnetic
field, and thus the topological phase region is refined as a closed
triangle in the phase diagram with chemical potential vs. Zeeman energy
(which is obviously different from the open hyperbolic region known
before). This prediction is confirmed again by an exact calculation
of quantum transport, where the zero bias peak of $2e^{2}/h$ in the
differential conductance spectrum, as the necessary evidence for the
Majorana zero modes, disappears when the magnetic field grows too
strong. For illustrations with practical hybrid systems, in the InSb
nanowire coupled to NbTiN, the accessible magnetic field range is
around 0.1 \textminus{} 1.5 T; when coupled to aluminum shell, the
accessible magnetic field range should be smaller than 0.12 T. These
predictions obviously clarify the current controversial issues about
some experiments of Majorana zero modes with hybrid nonawire.
\end{abstract}
\maketitle
\begin{figure}
\includegraphics[width=0.75\columnwidth]{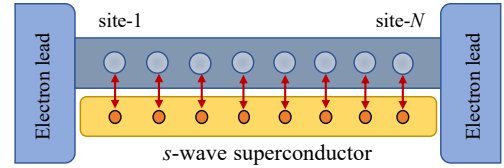}

\caption{Setup of the hybrid nanowire: The semiconductor nanowire is contacted
with the surface of the s-wave SC through tunneling. Through the virtual
exchanges of the quasi-excitations in the SC, this proximity effect
induces an effective pairing among the electrons in the nanowire.
In the experiment of quantum transport to probe Majorana zero modes,
the hybrid system is connected to the electron leads.}

\label{fig-hybrid}
\end{figure}

\noindent\textbf{Introduction} - The experiments for the Majorana
zero modes (MZMs) have attracted extensive attentions in recent years
\citep{kitaev_unpaired_2001,sau_generic_2010,alicea_new_2012,lutchyn_majorana_2018,reeg_transport_2017,stanescu_proximity-induced_2017,liu_andreev_2017},
for their novel fractional statistics and potential applications in
quantum computation \citep{kitaev_unpaired_2001,kitaev_fault-tolerant_2003,nayak_non-abelian_2008}.
In particular, the possible conductance signature for MZMs has been
shown in hybrid semiconductor-superconductor (HSS) systems \citep{mourik_signatures_2012,das_zero-bias_2012},
where a semiconductor nanowire with appreciable spin-orbit coupling
is contacted with an \emph{s}-wave superconductor (SC) providing the
SC proximity effect {[}see Fig. \ref{fig-hybrid}{]}. Under a proper
magnetic field, effectively a \emph{p}-wave pairing could be induced
in the hybrid nanowire, which gives birth to the MZMs localized at
the two ends of the open wire. In transport experiments, the existence
of MZMs would result in a zero bias peak (ZBP) in the differential
conductance spectrum, and the height of the ZBP should be $2e^{2}/h$
in the idealistic case at the zero temperature \citep{law_majorana_2009,flensberg_tunneling_2010}.

However, most of the current ZBP detections for MZMs do not reach
the idealistic height of $2e^{2}/h$ but lower \citep{mourik_signatures_2012,das_zero-bias_2012,nichele_scaling_2017,liu_universal_2021}.
As a result, such a ZBP signature alone cannot sufficiently confirm
the existence of MZMs as the unique reason apart from other possible
physical effects, such as Andreev bound states \citep{liu_andreev_2012,deng_majorana_2016,liu_andreev_2017,millo_what_2018},
and the Kondo effect \citep{sasaki_kondo_2000,lee_zero-bias_2012}.
To help narrow down the searching range in experiments, a more precise
phase diagram for different parameter regimes is urgently needed to
find the MZMs, for example, what is the valid extent for the magnetic
field where the MZMs could exist.

In previous investigations, the SC proximity effect was assumed to
be an \emph{s}-wave pairing in the nanowire with the pairing strength,
which is approximated as a constant \citep{fu_superconducting_2008,oreg_helical_2010,alicea_majorana_2010,sau_non-abelian_2010,stoudenmire_interaction_2011,sau_generic_2010},
and this is consistent with the result from the Bogoliubov-de-Gennes equation
projected in the low energy band \citep{stanescu_proximity_2010,sau_robustness_2010,potter_engineering_2011,reeg_transport_2017,stanescu_proximity-induced_2017,peng_strong_2015,reeg_finite-size_2017}.
It follows from this over-approximation that the topological region
bearing MZMs fills up the whole upper half of a hyperbolic curve in
the $\mu\text{-}B$ diagram ($\mu$, $B$ are the chemical potential
and Zeeman splitting of the nanowire respectively). Namely, these
approximated results indicate MZMs always exist no matter how strong
the magnetic field grows. In this letter, however, we find that indeed
this conclusion is not proper, and MZMs emerge only when the magnetic
field lies within a modest regime.

First, we utilize the Fr\"ohlich-Nakajima (Schrieffer-Wolff) transformation
to obtain an effective Hamiltonian for the nanowire, which encloses
the SC proximity effect and then gives an effective Kitaev model in
the HSS nanowire \citep{frohlich_theory_1950,nakajima_perturbation_1955,schrieffer_relation_1966}.
It turns out the effective pairing strength exhibits significant dependence
on the strong magnetic field. As a result, the topological region
bearing MZMs turns out to be a closed triangle in the $\mu\text{-}B$
diagram in comparison with the hyperbolic curve known before. Namely,
when the magnetic field is too strong, the system enters the non-topological
region, and thus MZMs disappear; when the magnetic field is weak,
this triangle region just returns the previous hyperbolic curve. It
is predicted from this phase diagram that the MZMs only emerge when
the magnetic field strength lies within a modest regime. For the current
experiments with InSb-NbTiN and InSb-Al, the valid magnetic field
ranges are $0.1-1.5\,\text{T}$ and $0.013-0.12\,\mathrm{T}$ respectively,
the latter of which is out of the range used in some of the current
experiments.

To confirm the above prediction based on the effective theory, we
further make an exact calculation on the differential conductance
in the transport measurement of the HSS nanowire by a quantum Langevin
equation \citep{dhar_quantum_2003,roy_majorana_2012,yang_master_2013,li_probing_2014}.
When the magnetic field increases from zero, a ZBP with $2e^{2}/h$
appear at a certain field strength, keeps for a while, and then disappears
at a certain higher strength, which is just consistent with the conclusion
from the effective Hamiltonian. This result provides a more precise
phase diagram which refines the searching region of MZMs in experiments.

\vspace{0.6em}\noindent\textbf{Effective Kitaev model in low energy
scale} - In this HSS system, the semiconductor nanowire is described
by the Hamiltonian \citep{alicea_new_2012}
\begin{equation}
\hat{H}_{\text{w}}=\int\mathrm{d}x\,\hat{\boldsymbol{\psi}}^{\dagger}(x)\Big[-\frac{\partial_{x}^{2}}{2m_{\mathrm{w}}}-\mu-i\alpha\sigma^{y}\partial_{x}+B\sigma^{z}\Big]\hat{\boldsymbol{\psi}}(x),\label{eq:H-x}
\end{equation}
where $\hat{\boldsymbol{\psi}}(x):=[\hat{\psi}_{\uparrow}(x),\hat{\psi}_{\downarrow}(x)]^{T}$,
and $\sigma^{y,z}$ are the Pauli matrices. Here, $\uparrow,\downarrow$
indicate the electron spins, $\alpha$ the spin-orbit coupling strength,
$m_{\mathrm{w}}$ the effective mass, $\mu$ the chemical potential
of the nanowire, and $B$ the Zeeman splitting from the external magnetic
field respectively.

The semiconductor nanowire is placed in contact with an \emph{s}-wave
SC providing the SC proximity effect, which is described by the BCS
Hamiltonian
\begin{equation}
\hat{H}_{\mathrm{sc}}=\sum_{\mathbf{k}}\epsilon_{\mathbf{k}}^{\mathrm{sc}}(\hat{c}_{\mathbf{k}\uparrow}^{\dagger}\hat{c}_{\mathbf{k}\uparrow}-\hat{c}_{-\mathbf{k}\downarrow}\hat{c}_{-\mathbf{k}\downarrow}^{\dagger})+\Delta_{\mathrm{s}}(\hat{c}_{\mathbf{k}\uparrow}^{\dagger}\hat{c}_{-\mathbf{k}\downarrow}^{\dagger}+\mathrm{H.c.}),\label{eq:H-sc}
\end{equation}
with $\epsilon_{\mathbf{k}}^{\mathrm{sc}}\equiv\mathbf{k}^{2}/2m_{\mathrm{sc}}-\mu_{\mathrm{sc}}$.
Hereafter the chemical potential of the \emph{s}-wave SC is set as
$\mu_{\mathrm{sc}}\equiv0$. The whole nanowire is contacted with
the surface of the \emph{s}-wave SC through the tunneling term
\begin{align}
\hat{H}_{\text{w-sc}} & =-\mathtt{J}_{\mathrm{s}}\sum_{s=\uparrow,\downarrow}\int\mathrm{d}x\,[\hat{\psi}_{s}^{\dagger}(x)\hat{c}_{s}(x,0,0)+\mathrm{H.c.}],\label{eq:Hw-sc}
\end{align}
where $\mathtt{J}_{\mathrm{s}}$ is the tunneling strength, $\hat{c}_{\mathbf{k}s}$
and $\hat{c}_{s}(\mathbf{x})$ are the Fourier images with each other.
The tunneling coupling induces an effective pairing among the electrons
in the nanowire and this proximity effect is caused by the virtual
exchanges of the quasi-excitations in the SC.

To describe the above mentioned virtual progress governed by the total
Hamiltonian $\hat{\mathcal{H}}\equiv\hat{H}_{\text{w}}+\hat{H}_{\mathrm{sc}}+\hat{H}_{\text{w-sc}}$,
we apply the Fr\"ohlich-Nakajima (Schrieffer-Wolff) transformation
to eliminate the degrees of freedom of the \emph{s}-wave SC \citep{frohlich_theory_1950,nakajima_perturbation_1955,schrieffer_relation_1966}.
When the coupling between the nanowire and the \emph{s}-wave SC is
weak enough, the effective Hamiltonian for the nanowire is obtained
as (see supplemental materials)
\begin{align}
 & \hat{H}_{\mathrm{eff}}=\int\frac{\mathrm{d}k}{2\pi}\,\Big\{(\tilde{\epsilon}_{\mathrm{w},k}-\tilde{\mu}_{k})\big[\hat{\varphi}_{\uparrow}^{\dagger}(k)\hat{\varphi}_{\uparrow}(k)+\hat{\varphi}_{\downarrow}^{\dagger}(k)\hat{\varphi}_{\downarrow}(k)\big]\nonumber \\
 & +i\tilde{\alpha}_{k}k\big[\hat{\varphi}_{\downarrow}^{\dagger}(k)\hat{\varphi}_{\uparrow}(k)-\mathrm{H.c.}\big]+\tilde{B}_{k}\big[\hat{\varphi}_{\uparrow}^{\dagger}(k)\hat{\varphi}_{\uparrow}(k)-\hat{\varphi}_{\downarrow}^{\dagger}(k)\hat{\varphi}_{\downarrow}(k)\big]\nonumber \\
 & +\tilde{\Delta}_{k}\big[\hat{\varphi}_{\uparrow}^{\dagger}(k)\hat{\varphi}_{\downarrow}^{\dagger}(-k)+\mathrm{H.c.}\big]\Big\}.\label{eq:Heff}
\end{align}
Here $\hat{\varphi}_{s}(k)$ is the Fourier transform of $\hat{\psi}_{s}(x)$.
$\tilde{\Delta}_{k}$ is the effective pairing strength induced by
the SC proximity effect, $\tilde{\epsilon}_{\mathrm{w},k}$, $\tilde{\mu}_{k}$,
$\tilde{\alpha}_{k}$, and $\tilde{B}_{k}$ are the corrected kinetic
energy ($\epsilon_{\mathrm{w},k}\equiv k^{2}/2m_{\mathrm{w}}$), chemical
potential, spin-orbit coupling and Zeeman splitting of the nanowire
respectively, i.e.,
\begin{align}
\tilde{\Delta}_{k} & =\Upsilon_{\mathrm{s}}\Big[1-\frac{\alpha^{2}k^{2}+B^{2}}{\Delta_{\mathrm{s}}^{2}}\Big]^{-\frac{1}{2}},\nonumber \\
\frac{\tilde{\epsilon}_{\mathrm{w},k}}{\epsilon_{\mathrm{w},k}} & =\frac{\tilde{\mu}_{k}}{\mu}=\frac{\tilde{\alpha}_{k}}{\alpha}=\frac{\tilde{B}_{k}}{B}=1-\frac{\tilde{\Delta}_{k}}{\Delta_{\mathrm{s}}}.\label{eq:Corrected}
\end{align}
Here, $\Upsilon_{\mathrm{s}}:=\mathtt{J}_{\mathrm{s}}^{2}\rho_{\mathrm{s}}$
describes the coupling strength between the nanowire and the s-wave
SC, with $\rho_{\mathrm{s}}$ as the density of states from the s-wave
SC, and approximately $\Upsilon_{\mathrm{s}}$ is a constant. Notice
that here the dependence on the magnetic field $B$ is well kept in
the above corrected parameters.

\begin{figure}
\includegraphics[width=1\columnwidth]{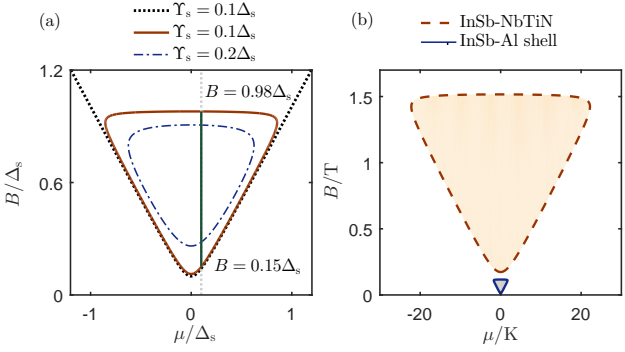}

\caption{(a) The topological phase diagram for HSS nanowire given by the effective
Hamiltonian (\ref{eq:Heff}) scaled by $\Delta_{\mathrm{s}}$. The
topological phase region bearing MZMs is a closed triangle. The dotted
black line is given by the approximated topological criterion $B^{2}-\mu^{2}=\Upsilon_{\mathrm{s}}^{2}$
\citep{ghosh_non-abelian_2010,alicea_new_2012}. The vertical line
is $\mu=0.1\,\Delta_{\mathrm{s}}$, and the valid range for the magnetic
field is $B\sim0.15-0.98\,\Delta_{\mathrm{s}}$. (b) The rescaled
phased diagram. For InSb nanowire (Land\'e factor $g\simeq50$) coupled
to NbTiN as the \emph{s}-wave SC ($\Delta_{\mathrm{s}}\simeq26\,\mathrm{K}$
\citep{mourik_signatures_2012}), the valid topological phase lies
in the yellow region, with the magnetic field range $B\sim0.15-1.5\,\mathrm{T}$;
for InSb nanowire coupled to aluminum shell ($\Delta_{\mathrm{s}}\simeq2\,\mathrm{K}$
\citep{lutchyn_majorana_2018,court_energy_2007}), the valid topological
phase lies in the smaller gray region, with the magnetic field range
around $0.012-0.12\,\mathrm{T}$, where MZMs could exist.}

\label{fig-phase}
\end{figure}

Under the open boundary condition, the effective Hamiltonian (\ref{eq:Heff})
has two edge modes localized at the two ends of the nanowire, whose
mode energies are zero, and they are just the MZMs. It can be proved
that the existence condition for the MZMs is given by the critical
condition $[\tilde{B}_{k}^{2}-\tilde{\mu}_{k}^{2}-\tilde{\Delta}_{k}^{2}]\big|_{k=0}=0$
(see supplemental materials), which determines the topological phase
region \citep{ghosh_non-abelian_2010,alicea_new_2012}. It turns out
this topological region bearing MZMs appears as a closed triangle
in the $\mu\text{-}B$ phase diagram (Fig. \ref{fig-phase}). For
a fixed chemical potential $\mu$, the MZMs could emerge only if the
magnetic field strength must properly lie in a modest range. When
the magnetic field strength exceeds the range of the magnetic field
determined by refined phase region, the effective Hamiltonian (\ref{eq:Heff})
of the hybrid nanowire does not support the existence of the MZMs\footnote{With
the increase of the magnetic field, the \emph{s}-wave SC gap $\Delta_{\mathrm{s}}$
would also decrease. Here this effect is not considered, and $\Delta_{\mathrm{s}}$
is treated as a constant independent of the magnetic field. If this
effect is considered, the topological region in Fig.\,\ref{fig-phase}
would be smaller.}.

For the weak field situation ($B\ll\Delta_{\mathrm{s}}$), in the
low energy regime ($k\simeq0$), the induced paring strength can be
approximated as a constant $\tilde{\Delta}_{k}\simeq\Upsilon_{\mathrm{s}}$
{[}see Eq. (\ref{eq:Corrected}){]}. Correspondingly, the above topological
phase condition is reduced as $B^{2}-\mu^{2}=\Upsilon_{\mathrm{s}}^{2}/(1-\Upsilon_{\mathrm{s}}/\Delta_{\mathrm{s}})^{2}\simeq\Upsilon_{\mathrm{s}}^{2}$,
which just returns the hyperbolic curve sufficiently studied in previous
literatures \citep{ghosh_non-abelian_2010,alicea_new_2012}. Indeed
the bottom part of the close triangle region and the hyperbolic curve
fit well with each other (Fig. \ref{fig-phase}), which is consistent
with the fact that the hyperbolic curve comes from an effective theory
in the low energy regime.

To have a more clear understanding on above observations, we consider
the above effective Hamiltonian in a new representation {[}by taking
the first three bracket terms of (\ref{eq:Heff}) into diagonalization{]},
\begin{align}
\hat{H}_{\mathrm{eff}}= & \int\frac{\mathrm{d}k}{2\pi}\,\Big\{\tilde{\varepsilon}_{k+}\hat{\varphi}_{+}^{\dagger}(k)\hat{\varphi}_{+}(k)+\tilde{\varepsilon}_{k-}\hat{\varphi}_{-}^{\dagger}(k)\hat{\varphi}_{-}(k)\nonumber \\
+ & \frac{1}{2}\tilde{\Delta}_{k}^{(\mathrm{p})}\big[\hat{\varphi}_{+}^{\dagger}(k)\hat{\varphi}_{+}^{\dagger}(-k)+\hat{\varphi}_{-}^{\dagger}(k)\hat{\varphi}_{-}^{\dagger}(-k)+\mathrm{H.c.}\big]\nonumber \\
+ & \tilde{\Delta}_{k}^{(\mathrm{s})}\big[\hat{\varphi}_{+}^{\dagger}(k)\hat{\varphi}_{-}^{\dagger}(-k)+\mathrm{H.c.}\big]\Big\},
\end{align}
which appears as an effective Kitaev model with spins \citep{kitaev_unpaired_2001,potter_majorana_2011}.
Here, $\tilde{\varepsilon}_{k\pm}=\tilde{\epsilon}_{\mathrm{w},k}-\tilde{\mu}_{k}\pm\sqrt{\tilde{B}_{k}^{2}+\tilde{\alpha}_{k}^{2}k^{2}}$,
and
\begin{gather}
\tilde{\Delta}_{k}^{(\mathrm{p})}:=\frac{\alpha k\tilde{\Delta}_{k}}{\sqrt{B^{2}+\alpha^{2}k^{2}}},\quad\tilde{\Delta}_{k}^{(\mathrm{s})}:=\frac{B\tilde{\Delta}_{k}}{\sqrt{B^{2}+\alpha^{2}k^{2}}},\nonumber \\
\left[\begin{array}{c}
\hat{\varphi}_{+}(k)\\
\hat{\varphi}_{-}(k)
\end{array}\right]:=\left[\begin{array}{cc}
\cos\vartheta_{k} & i\sin\vartheta_{k}\\
-i\sin\vartheta_{k} & \cos\vartheta_{k}
\end{array}\right]\cdot\left[\begin{array}{c}
\hat{\varphi}_{\uparrow}(k)\\
\hat{\varphi}_{\downarrow}(k)
\end{array}\right],
\end{gather}
with $\tan2\vartheta_{k}=\tilde{\alpha}_{k}k/\tilde{B}_{k}=\alpha k/B$.

In the above representation, $\tilde{\Delta}_{k}^{(\mathrm{s})}$
and $\tilde{\Delta}_{k}^{(\mathrm{p})}$ are effectively regarded
as the \emph{s}-wave and \emph{p}-wave pairing strength respectively.
With the increase of the magnetic field $B$, the \emph{p}-wave pairing
$\tilde{\Delta}_{k}^{(\mathrm{p})}$ becomes weaker and weaker, while
relatively the \emph{s}-wave pairing $\tilde{\Delta}_{k}^{(\mathrm{s})}$
becomes more dominative. Thus when the magnetic field is too strong,
the nanowire system enters the non-topological phase region, and the
MZMs would disappear. Indeed, the disappearance of ZBP in strong magnetic
field was claimed in some of the current experiment signatures \citep{mourik_signatures_2012,das_zero-bias_2012},
which is well explained by our current predictions.

\vspace{0.6em}\noindent\textbf{Quantum transport} - To confirm the
above observations from the effective Hamiltonian, here we make an
exact calculation on the transport behavior of the hybrid nanowire.
To this end, the above continuous Hamiltonian (\ref{eq:H-x}) is firstly
discretized with respect to an $N$-site system as \citep{stoudenmire_interaction_2011,liu_zero-bias_2012}
\begin{align}
\hat{H}_{\mathrm{w}}= & \sum_{n,s}-\frac{J}{2}(\hat{d}_{n,s}^{\dagger}\hat{d}_{n+1,s}+\hat{d}_{n+1,s}^{\dagger}\hat{d}_{n,s})-(\mu-J)\hat{d}_{n,s}^{\dagger}\hat{d}_{n,s}\nonumber \\
 & +\sum_{n}\frac{\alpha}{2}(\hat{d}_{n,\downarrow}^{\dagger}\hat{d}_{n+1,\uparrow}-\hat{d}_{n,\uparrow}^{\dagger}\hat{d}_{n+1,\downarrow}+\mathrm{H.c.})\nonumber \\
 & +\sum_{n}B(\hat{d}_{n,\uparrow}^{\dagger}\hat{d}_{n,\uparrow}-\hat{d}_{n,\downarrow}^{\dagger}\hat{d}_{n,\downarrow}).\label{eq:H-w}
\end{align}
Correspondingly, the tunneling term (\ref{eq:Hw-sc}) between the
nanowire and the \emph{s}-wave SC becomes $\hat{H}_{\text{w-sc}}=-\sum_{n,\mathbf{k}s}(\mathtt{J}_{n,\mathbf{k}s}\,\hat{d}_{n,s}^{\dagger}\hat{c}_{\mathbf{k}s}+\mathrm{H.c.})$,
with the tunneling strength $|\mathtt{J}_{m,\mathbf{k}s}|=|\mathtt{J}_{n,\mathbf{k}s'}|:=\mathtt{J}_{\mathbf{k}}$
for $m\neq n$, $s\neq s'$. We consider two electron leads are contacted
with the two ends of the nanowire, which are described by $\hat{H}_{\text{e-}x}=\sum_{\mathbf{k}s}\,\varepsilon_{x,\mathbf{k}}\,\hat{b}_{x,\mathbf{k}s}^{\dagger}\hat{b}_{x,\mathbf{k}s}$,
with $x=1,N$ as the site number contacting with the electron lead.
The tunneling interaction between the nanowire and lead-$x$ is $\hat{H}_{\text{w-}x}=-\sum_{\mathbf{k}s}(\mathtt{g}_{x,\mathbf{k}}\,\hat{d}_{x,s}^{\dagger}\hat{b}_{x,\mathbf{k}s}+\mathrm{H.c.})$.

Notice that, similar to the two electron leads, indeed the \emph{s}-wave
SC also can be regarded as the third fermionic bath interacting with
the nanowire. Thus, it follows from the Heisenberg equations of $\hat{d}_{ns}$
(nanowire), $\hat{c}_{\mathbf{k}s}$ (\emph{s}-wave SC), and $\hat{b}_{x,\mathbf{k}s}$
(electron leads) that a quantum Langevin equation is derived to describe
the transport dynamics (supplemental materials) as \citep{dhar_quantum_2003,roy_majorana_2012,yang_master_2013,li_probing_2014}
\begin{equation}
\partial_{t}\hat{\mathbf{d}}=-i\mathbf{H}_{\mathrm{w}}\cdot\hat{\mathbf{d}}-\int_{0}^{t}\mathrm{d}\tau\,\mathbf{D}(t-\tau)\cdot\hat{\mathbf{d}}(\tau)+i\hat{\boldsymbol{\xi}}_{\mathrm{sc}}+i\hat{\boldsymbol{\xi}}_{\mathrm{e}}.\label{eq:Langevin}
\end{equation}
Here, $\hat{\mathbf{d}}(t):=(\hat{\boldsymbol{d}}_{1},\dots,\hat{\boldsymbol{d}}_{N})^{T}$
is a $4N$-vector form with $N$ blocks $\hat{\boldsymbol{d}}_{n}:=(\hat{d}_{n\uparrow},\hat{d}_{n\downarrow},\,\hat{d}_{n\uparrow}^{\dagger},\hat{d}_{n\downarrow}^{\dagger})^{T}$.
Then the nanowire Hamiltonian (\ref{eq:H-w}) is rewritten as $\hat{H}_{\mathrm{w}}\equiv\frac{1}{2}\hat{\mathbf{d}}^{\dagger}\cdot\mathbf{H}_{\mathrm{w}}\cdot\hat{\mathbf{d}}$,
with a $4N\times4N$ matrix $\mathbf{H}_{\mathrm{w}}$. The dissipation
kernel $\mathbf{D}(t)\equiv\mathbf{D}_{\mathrm{e}}(t)+\mathbf{D}_{\mathrm{sc}}(t)$
contains the contributions from both the two electron leads and the
\emph{s}-wave SC, and $\hat{\boldsymbol{\xi}}_{\mathrm{e}}(t)$ and
$\hat{\boldsymbol{\xi}}_{\mathrm{sc}}(t)$ are the corresponding random
forces respectively.

This Langevin equation (\ref{eq:Langevin}) of $\hat{\mathbf{d}}(t)$
is exactly solvable in the Fourier space, namely,
\begin{equation}
\tilde{\mathbf{d}}(\omega)=\mathbf{G}(\omega)\cdot\big[\hat{\mathbf{d}}_{(t=0)}+i\tilde{\boldsymbol{\xi}}_{\mathrm{sc}}(\omega)+i\tilde{\boldsymbol{\xi}}_{\mathrm{e}}(\omega)\big],
\end{equation}
where $[\mathbf{G}(\omega)]_{4N\times4N}$ is the Green function of
the nanowire,
\begin{align}
\mathbf{G}(\omega) & =i\big[\omega^{+}-\mathbf{H}_{\mathrm{w}}+i\tilde{\mathbf{D}}_{\mathrm{sc}}(\omega)+i\tilde{\mathbf{D}}_{\mathrm{e}}(\omega)\big]^{-1}\nonumber \\
 & \equiv i\Big\{\omega^{+}-\mathbf{H}_{\mathrm{w}}-\tilde{\mathbf{V}}_{\mathrm{s}}(\omega)+\frac{i}{2}[\tilde{\boldsymbol{\Gamma}}_{\mathrm{s}}(\omega)+\tilde{\boldsymbol{\Gamma}}_{\mathrm{e}}(\omega)]\Big\}^{-1},\label{eq:Green}
\end{align}
with $\omega^{+}\equiv\omega+i\epsilon$ ($\epsilon$ is an infinitesimal).
Here $\tilde{\mathbf{D}}_{\mathrm{sc}}(\omega)\equiv\tilde{\boldsymbol{\Gamma}}_{\mathrm{s}}(\omega)/2+i\tilde{\mathbf{V}}_{\mathrm{s}}(\omega)$
is the Fourier image of the dissipation kernel $\mathbf{D}_{\mathrm{sc}}(t)$
from the \emph{s}-wave SC, where the ``real part'' $\tilde{\boldsymbol{\Gamma}}_{\mathrm{s}}(\omega)$
leads to the system dissipation, and the ``imaginary part'' $\tilde{\mathbf{V}}_{\mathrm{s}}(\omega)$
provides an effective interaction to the nanowire Hamiltonian.

Specifically, $\tilde{\mathbf{D}}_{\mathrm{sc}}(\omega):=\mathrm{diag}\{\tilde{\mathsf{D}}_{\mathrm{s}},\dots,\tilde{\mathsf{D}}_{\mathrm{s}}\}$
is a block-diagonal matrix, with blocks $\tilde{\mathsf{D}}_{\mathrm{s}}(\omega):=\tilde{\mathsf{\Gamma}}_{\mathrm{s}}(\omega)/2+i\tilde{\mathsf{V}}_{\mathrm{s}}(\omega)$,
where
\begin{align}
\tilde{\mathsf{V}}_{\mathrm{s}}(\omega) & :=-\frac{\Theta(\Delta_{\mathrm{s}}-|\omega|)\Upsilon_{\mathrm{s}}}{\sqrt{\Delta_{\mathrm{s}}^{2}-\omega^{2}}}\left[\begin{array}{cccc}
\omega &  &  & -\Delta_{\mathrm{s}}\\
 & \omega & \Delta_{\mathrm{s}}\\
 & \Delta_{\mathrm{s}} & \omega\\
-\Delta_{\mathrm{s}} &  &  & \omega
\end{array}\right],\nonumber \\
\tilde{\mathsf{\Gamma}}_{\mathrm{s}}(\omega) & :=\frac{2\Theta(|\omega|-\Delta_{\mathrm{s}})\Upsilon_{\mathrm{s}}}{\sqrt{\omega^{2}-\Delta_{\mathrm{s}}^{2}}}\,|\omega|\mathbf{1}_{4\times4}.
\end{align}
 Here, $\Upsilon_{\mathrm{s}}(\omega):=\pi\sum_{\mathbf{k}}|\mathtt{J}_{\mathbf{k}}|^{2}\delta(\omega-\epsilon_{\mathbf{k}}^{\mathrm{sc}})\rightarrow\pi|\mathtt{J}_{\mathrm{s}}(\omega)|^{2}\rho_{\mathrm{s}}(\omega)$
is introduced as the spectral density of the coupling with the \emph{s}-wave
SC, which is approximated as a constant coupling strength $\Upsilon_{\mathrm{s}}(\omega)\simeq\Upsilon_{\mathrm{s}}$
{[}this notion is just consistent with the one in the above continuous
situation (\ref{eq:Corrected}){]}.

The dissipation kernels of the two electron leads also give $\tilde{\mathbf{D}}_{\mathrm{e}}(\omega)\equiv\tilde{\boldsymbol{\Gamma}}_{\mathrm{e}}(\omega)/2+i\tilde{\mathbf{V}}_{\mathrm{e}}(\omega)$,
while $\tilde{\mathbf{V}}_{\mathrm{e}}(\omega)\simeq0$ in usual transport
experiments, only with $\tilde{\boldsymbol{\Gamma}}_{\mathrm{e}}(\omega):=\boldsymbol{\Gamma}_{1}+\boldsymbol{\Gamma}_{N}$
left providing dissipation to the system. Here $\boldsymbol{\Gamma}_{1}:=\mathrm{diag}\{\mathsf{\Gamma}_{1},\mathbf{0},\dots,\mathbf{0}\}$
and $\boldsymbol{\Gamma}_{N}:=\mathrm{diag}\{\mathbf{0},\dots,\mathbf{0},\mathsf{\Gamma}_{N}\}$
are the dissipation matrices from the two electron leads respectively,
where $\mathsf{\Gamma}_{x}:=\Upsilon_{x}\,\mathbf{1}_{4\times4}$
($x=1,N$), and $\Upsilon_{x}$ indicates the coupling strength with
lead-$x$, defined from the coupling spectral density $\Upsilon_{x}(\omega):=2\pi\sum_{\mathbf{k}}|\mathtt{g}_{x,\mathbf{k}}|^{2}\delta(\omega-\varepsilon_{x,\mathbf{k}})\simeq\Upsilon_{x}$.

It is worth noting that, without deriving the effective Hamiltonian
in priori with any approximations, the SC proximity effect is naturally
presented as the off-diagonal elements of $\tilde{\mathsf{V}}_{\mathrm{s}}(\omega)$
in the dynamical propagator (\ref{eq:Green}), which just indicates
the onsite \emph{s}-wave pairing for the nanowire \citep{liu_zero-bias_2012,sau_robustness_2010,potter_majorana_2011,liu_andreev_2017}.
Moreover, a Heaviside function naturally appears in both $\tilde{\mathsf{V}}_{\mathrm{s}}(\omega)$
and $\tilde{\mathsf{\Gamma}}_{\mathrm{s}}(\omega)$, which indicates
a complementary effect of the SC proximity: when the system energy
scale lies within the \emph{s}-wave gap $|\omega|<\Delta_{\mathrm{s}}$,
the \emph{s}-wave SC just provides the effective pairing interaction
without any dissipation; in contrast, outside the gap $|\omega|>\Delta_{\mathrm{s}}$,
the SC proximity does not give the effective pairing, but only brings
in the dissipation effect similarly as the normal leads.

\begin{figure}
\includegraphics[width=1\columnwidth]{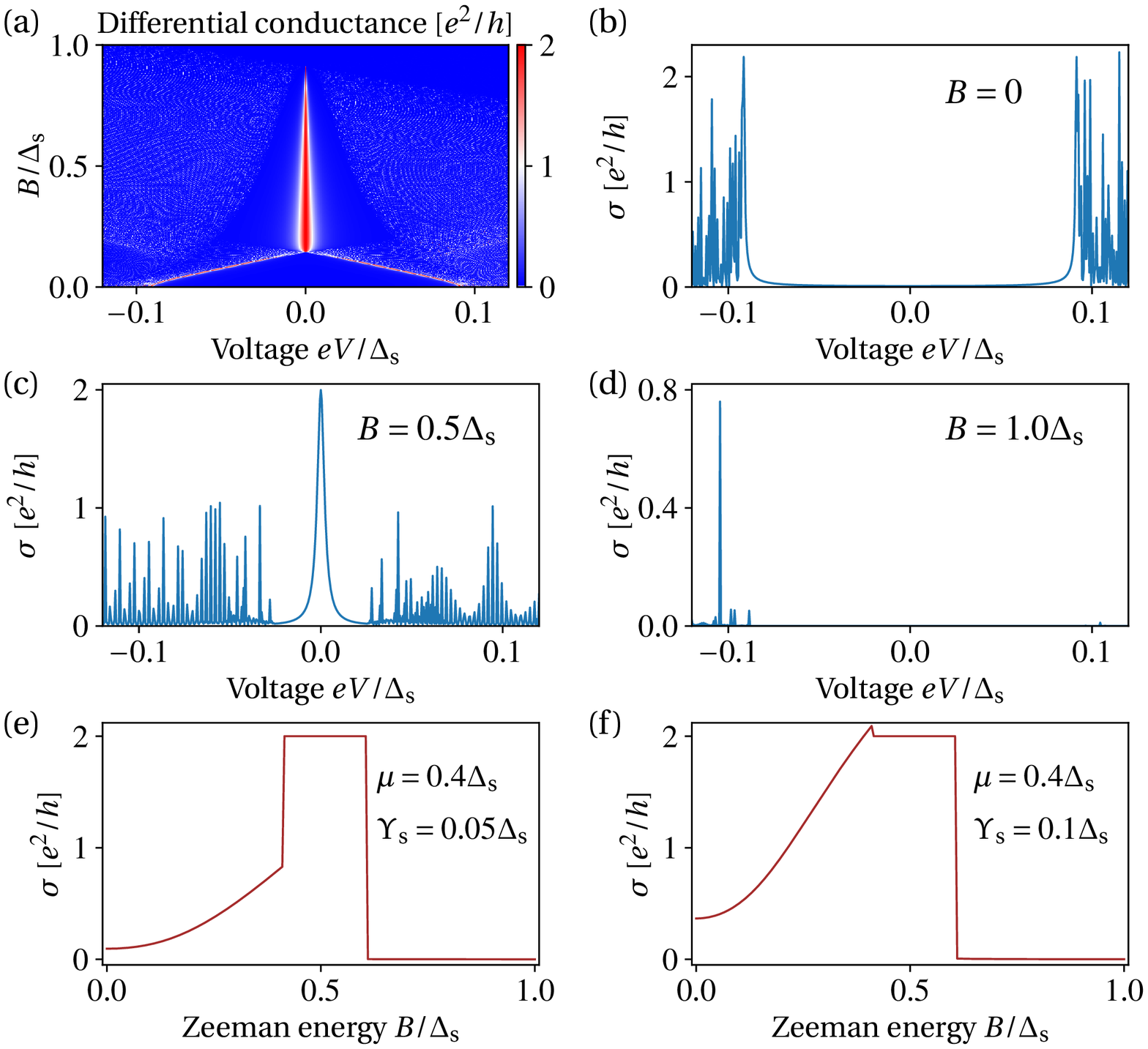}

\caption{(a) The differential conductance $\sigma=dI_{1}/dV$ depending on
the bias voltage and the magnetic field (site number $N=500$). Here
$\Delta_{\mathrm{s}}$ is set as the energy unit, and the other parameters
are set as $\mu=0.1\Delta_{\mathrm{\mathrm{s}}}$, $\alpha=0.15\Delta_{\mathrm{\mathrm{s}}}$,
$J=0.5\Delta_{\mathrm{\mathrm{s}}}$, $\Upsilon_{\mathrm{s}}=0.1\Delta_{\mathrm{\mathrm{s}}}$,
$\Upsilon_{1}=\Upsilon_{N}=0.025\Delta_{\mathrm{\mathrm{s}}}$. (b,
c, d) The differential conductance depending on the bias voltage,
when the Zeeman energies are fixed as $B/\Delta_{\mathrm{s}}=0,\,0.5,\,1.0$
respectively. (e, f) The differential conductance at $V=0$ changing
with the magnetic field. Under certain parameters, the ZBP could be
higher than $2e^{2}/h$, which indicates this is not from the MZMs.}

\label{fig-didv}
\end{figure}

\vspace{0.6em}\noindent\textbf{Differential conductance} - To study
the transport current, we consider the initial states of the three
baths (the two normal leads, and the \emph{s}-wave SC) are in the
Fermi-Dirac distributions at the zero temperature. The chemical potentials
of the \emph{s}-wave SC and the electron lead-$N$ are set as $\mu_{\mathrm{sc}}=\mu_{N}=0$,
while the lead-1 is $\mu_{1}=eV$ with $V$ as the bias voltage.

The electric current flowing from lead-1 to the nanowire is obtained
from the changing rate of the total electron number in lead-1, i.e.,
$\hat{I}_{1}(t):=-e\,\partial_{t}\sum_{\mathbf{k}s}\langle\hat{b}_{1,\mathbf{k}s}^{\dagger}\hat{b}_{1,\mathbf{k}s}\rangle$.
After a long enough time relaxation $t\rightarrow\infty$, a steady
current is achieved. From the above Langevin equation, the differential
conductance $\sigma\equiv dI_{1}/dV$ is obtained as (see supplemental
materials) \citep{roy_majorana_2012,li_probing_2014}
\begin{align}
\sigma= & \frac{e^{2}}{h}\big\{\mathrm{tr}[\mathbf{G}^{\dagger}\boldsymbol{\Gamma}_{1}^{+}\mathbf{G}\boldsymbol{\Gamma}_{N}]_{(eV)}+\mathrm{tr}[\mathbf{G}^{\dagger}\boldsymbol{\Gamma}_{1}^{+}\mathbf{G}\tilde{\boldsymbol{\Gamma}}_{\mathrm{s}}]_{(eV)}\nonumber \\
 & +\mathrm{tr}[\mathbf{G}^{\dagger}\boldsymbol{\Gamma}_{1}^{+}\mathbf{G}\boldsymbol{\Gamma}_{1}^{-}]_{(eV)}+\mathrm{tr}[\mathbf{G}^{\dagger}\boldsymbol{\Gamma}_{1}^{+}\mathbf{G}\boldsymbol{\Gamma}_{1}^{-}]_{(-eV)}\big\}.\label{eq:didv}
\end{align}
Here the dissipation matrices $\boldsymbol{\Gamma}_{1,N}^{\pm}$ are
given by $\boldsymbol{\Gamma}_{1}^{\pm}:=\mathrm{diag}\{\mathsf{\Gamma}_{1}^{\pm},\mathbf{0},\dots,\mathbf{0}\}$,
and $\boldsymbol{\Gamma}_{N}^{\pm}:=\mathrm{diag}\{\mathbf{0},\dots,\mathbf{0},\mathsf{\Gamma}_{N}^{\pm}\}$,
with the $4\times4$ blocks $\mathsf{\Gamma}_{x}^{+}:=\Upsilon_{x}\,\mathrm{diag}\{1,1,0,0\}$,
and $\mathsf{\Gamma}_{x}^{-}:=\Upsilon_{x}\,\mathrm{diag}\{0,0,1,1\}$.

The first two terms in Eq. (\ref{eq:didv}) come from the electron
exchanges among lead-1 to lead-$N$ and the \emph{s}-wave SC respectively;
the last two terms indicate the contribution from the Andreev reflection
between lead-1 and the nanowire, which gives the ZBP of $2e^{2}/h$
as the necessary signature for the emergence of MZMs at the zero temperature
\citep{law_majorana_2009,flensberg_tunneling_2010,roy_majorana_2012,li_probing_2014}.
Up to now, no other approximations are made except the form of the
coupling spectral density, thus the obtained result is exact enough
even for the situations that the coupling strength or the magnetic
field is quite strong.

The numerical results for the differential conductance (\ref{eq:didv})
under different physical conditions are illustrated in Fig. \ref{fig-didv}.
It is shown that a ZBP with height $2e^{2}/h$ appears in the conductance
spectrum when the magnetic field $B$ properly lies in a continuous
regime of modest strength, and the ZBP of $2e^{2}/h$ does not represent
Majorana zero modes when the magnetic field strength exceeds the range
of the magnetic field determined by refined phase region {[}see Fig.
\ref{fig-phase}{]}. Then the ZBP disappears when the magnetic field
strength is too weak or too strong, thus, it is confirmed that the
MZMs do not exist in these regimes. This observation is just consistent
with the result obtained from the above effective Hamiltonian (Fig.
\ref{fig-phase}). The variation trend of ZBP signature with the magnetic
field is fit with the observed result in experiment \citep{mourik_signatures_2012,das_zero-bias_2012,zhang_quantized_2018}.

It is specially worth noting that under certain conditions here the
ZBP could be even higher than $2e^{2}/h$ {[}see Fig. \ref{fig-didv}(e,
f){]}. In finite temperatures, such a ZBP would be as low as $2e^{2}/h$
and thus might be confused with the signature from MZMs. Therefore
it should be emphasized that the ZBP of $2e^{2}/h$ is the necessary
but not sufficient condition for the MZMs.

\vspace{0.6em}\noindent\textbf{Summary} - By examining the the low-energy
effective model for the hybrid system with the semiconductor nanowire
in proximity to the \emph{s}-wave superconductor, we obtain a refined
topological phase diagram where the Majorana zero modes (MZMs) could
exist. The valid topological phase region bearing MZMs appears as
a closed triangle in the $\mu\text{-}B$ phase diagram, in comparison
with the open hyperbolic region known before. These predictions are
also confirmed by the exact calculation about the quantum transport
based on the quantum Langevin equation: in the transport spectrum,
the zero bias peak with $2e^{2}/h$, as the necessary signature for
MZMs, disappears when the magnetic field grows too strong. Therefore,
to search MZMs in this hybrid nanowire system, we suggest that the
magnetic field strength be properly set within a modest range according
to our novel phase diagram.

For the electron-doped InSb nanowire coupled to NbTiN ($\Delta_{\mathrm{s}}\simeq26\,\mathrm{K}$),
the chemical potential is around $\mu\sim0-10\,\mathrm{K}$, the spin-orbit
energy is around $U\equiv2m_{\mathrm{w}}\alpha^{2}\sim1-3\,\mathrm{K}$
\citep{alicea_majorana_2010,mourik_signatures_2012,das_zero-bias_2012,court_energy_2007,nadj-perge_spectroscopy_2012,hong_terahertz_2013,van_heck_conductance_2016,lutchyn_majorana_2018},
and the above results show that the proper range for the magnetic
field is around $B\sim0.1-1.5\,\mathrm{T}$ where MZMs could exist.
And for InSb nanowire coupled to aluminum shell ($\Delta_{\mathrm{s}}\simeq2\,\mathrm{K}$)
\citep{lutchyn_majorana_2018,court_energy_2007}, the convincing range
for the magnetic field is no greater than $0.12\,\mathrm{T}$. However,
some of the recent experiments claim the emergence of MZMs when the
magnetic field strength exceeds much beyond the valid range that we
predicted in this letter. It is believed that our current theoretical
study eventually solves the corresponding controversies about MZM
experiments.

\vspace{0.6em}

The authors appreciate quite much for the helpful discussion with
Y. Chen in CAEP, Y.-N. Fang in Yunnan University. This study is supported
by NSF of China (Grant No. 12088101 and No. 11905007), National Basic
Research Program of China (Grant No. 2016YFA0301201), NSAF (Grants
No. U1930403 and No. U1930402).

\bibliographystyle{apsrev4-2}
\bibliography{Refs}

\appendix
\begin{widetext}

\section{The low energy effective Hamiltonian of nanowire}

In this section, we derive the effective Hamiltonian of the nanowire
by using the Fr\"ohlich-Nakajima transformation. For the hybird semiconductor-superconductor
nanowire system, the total Hamiltonian has three basic terms: $\mathcal{\hat{H}}=\hat{H}_{\mathrm{w}}+\hat{H}_{\mathrm{sc}}+\hat{H}_{\mathrm{w\text{-}sc}}$.
The nanowire Hamiltonian in continuous model is
\begin{equation}
\hat{H}_{\mathrm{w}}=\int\mathrm{d}x\,\hat{\boldsymbol{\psi}}^{\dagger}(x)[-\frac{\partial_{x}^{2}}{2m_{\mathrm{w}}}-\mu-i\alpha\sigma^{y}\partial_{x}+B\sigma^{z}]\hat{\boldsymbol{\psi}}(x),\label{eq:nonawireH}
\end{equation}
where $\hat{\boldsymbol{\psi}}(x)=[\hat{\psi}_{\uparrow}(x),\hat{\psi}_{\downarrow}(x)]^{T}$,
and $\sigma^{y,z}$ are the Pauli matrices. Here, $\uparrow,\downarrow$
indicate the electron spins, $\alpha$ the spin-orbit coupling strength,
$m_{\mathrm{w}}$ the effective mass, $\mu$ the chemical potential
of the nanowire, and $B$ the Zeeman splitting from the external magnetic
field respectively. The \textsl{s}-wave superconductor (SC) providing
the SC proximity effect for the nanowire is described by
\begin{equation}
\hat{H}_{\mathrm{sc}}=\int\frac{\mathrm{d}^{3}k}{(2\pi)^{3}}\,\epsilon_{\mathbf{k}}^{\mathrm{sc}}(\hat{c}_{\uparrow}^{\dagger}(\mathbf{k})\hat{c}_{\uparrow}(\mathbf{k})-\hat{c}_{\downarrow}(-\mathbf{k})\hat{c}_{\downarrow}^{\dagger}(-\mathbf{k}))+\Delta_{\text{s}}(\hat{c}_{\uparrow}^{\dagger}(\mathbf{k})\hat{c}_{\downarrow}^{\dagger}(-\mathbf{k})+\mathrm{H.c.})\label{eq:sc}
\end{equation}
with the kinetic energy $\epsilon_{\mathbf{k}}^{\mathrm{sc}}$ and
real pairing potential $\Delta_{\mathrm{s}}$ of SC. The fermion operator
$\hat{c}_{\uparrow,\downarrow}(\mathbf{k})$ follows the anti-commutation
relation $\{\hat{c}_{s}(\mathbf{k}),\,\hat{c}_{s'}^{\dagger}(\mathbf{k}')\}_{+}=(2\pi)^{3}\delta_{ss'}\delta(\mathbf{k}-\mathbf{k}')$.
The tunneling interaction between the nanowire and \emph{s}-wave superconductor
is
\begin{equation}
\begin{aligned}\hat{H}_{\mathrm{w\text{-\ensuremath{\mathrm{sc}}}}}=-\mathtt{J}_{\mathrm{s}}\sum_{s}\int\mathrm{d}x\,[\hat{\psi}_{s}^{\dagger}(x)\hat{c}_{s}(x,0,0)+\hat{c}_{s}^{\dagger}(x,0,0)\hat{\psi}_{s}(x)].\end{aligned}
\label{eq:tunneling}
\end{equation}
Here, $\mathtt{J}_{\mathrm{s}}$ describes the tunneling strength
between the nanowire and the \emph{s}-wave SC, and $\hat{c}_{s}(\mathbf{x})$
is Fourier image of $\hat{c}_{s}(\mathbf{k})$:
\begin{equation}
\hat{c}_{s}(\mathbf{x})=\frac{1}{(2\pi)^{3}}\int\hat{c}_{s}(\mathbf{k})e^{i\mathbf{k}\cdot\mathbf{x}}\mathrm{d}^{3}k\label{eq:Fourier image}
\end{equation}
Similarly, the above nanowire Hamiltonian (\ref{eq:nonawireH}) and
tunneling Hamiltonian (\ref{eq:tunneling}) in Fourier space become
\begin{equation}
\begin{aligned}\hat{H}_{\mathrm{w}} & =\int\frac{\mathrm{d}k_{x}}{2\pi}\,\hat{\boldsymbol{\varphi}}^{\dagger}(k_{x})[\epsilon_{k_{x}}+\alpha k_{x}\sigma^{y}+B\sigma^{z}]\hat{\boldsymbol{\varphi}}(k_{x}),\\
\hat{H}_{\mathrm{w\text{-}sc}} & =-\mathtt{J}_{\mathrm{s}}\sum_{s}\int\frac{\mathrm{d}^{3}k}{(2\pi)^{3}}\,[\varphi_{s}^{\dagger}(k_{x})\hat{c}_{s}(\mathbf{k})+\hat{c}_{s}^{\dagger}(\mathbf{k})\varphi_{s}(k_{x})].
\end{aligned}
\label{eq:kH}
\end{equation}
Here, $\hat{\boldsymbol{\varphi}}(k_{x}\text{)}=[\hat{\varphi}_{\uparrow}(k_{x}\text{)},\hat{\varphi}_{\downarrow}(k_{x}\text{)}]^{T}$
is obtained by Fourier transform of two-component operator $\hat{\boldsymbol{\psi}}(x)$
in real space, $\epsilon_{k_{x}}\equiv\frac{k_{x}^{2}}{2m_{\mathrm{w}}}-\mu$
is the electron kinetic energy of the nanowire and $k_{x}$ is the
$x$ component of $\mathbf{k}$.

By the following Bogoliubov transformation
\begin{gather}
\hat{\eta}_{\uparrow}(\mathbf{k}):=\cos\theta_{\mathbf{k}}\hat{c}_{\uparrow}(\mathbf{k})+\sin\theta_{\mathbf{k}}\hat{c}_{\downarrow}^{\dagger}(-\mathbf{k}),\nonumber \\
\hat{\eta}_{\downarrow}^{\dagger}(-\mathbf{k}):=-\sin\theta_{\mathbf{k}}\hat{c}_{\uparrow}(\mathbf{k})+\cos\theta_{\mathbf{k}}\hat{c}_{\downarrow}^{\dagger}(-\mathbf{k}),\label{eq:bogoliubov transform}
\end{gather}
with $\tan2\theta_{\mathbf{k}}=\Delta_{\mathrm{s}}/\epsilon_{\mathbf{k}}^{\mathrm{sc}}$,
the Hamiltonian of the \emph{s}-wave SC can be diagonally reduced
to
\begin{gather}
\hat{H}_{\mathrm{sc}}=\int\frac{\mathrm{d}^{3}k}{(2\pi)^{3}}\,E_{\mathbf{k}}^{\mathrm{sc}}[\hat{\eta}_{\uparrow}^{\dagger}(\mathbf{k})\hat{\eta}_{\uparrow}(\mathbf{k})+\hat{\eta}_{\downarrow}^{\text{\ensuremath{\dagger}}}(-\mathbf{k})\hat{\eta}_{\downarrow}(-\mathbf{k})],\label{eq:Ksc}
\end{gather}
with the excitation spectrum of the SC $E_{\mathbf{k}}^{\mathrm{sc}}=\sqrt{[\epsilon_{\mathbf{k}}^{\mathrm{sc}}]^{2}+\Delta_{\text{s}}^{2}}$.
Then the tunneling Hamiltonian $\hat{H}_{\mathrm{w\text{-\ensuremath{\mathrm{sc}}}}}$
{[}Eq.\,(\ref{eq:tunneling}){]} is rewritten by the quasi-particle
operators $\hat{\eta}_{s}(\mathbf{k})$ as
\begin{equation}
\begin{aligned}\hat{H}_{\text{w-sc}}= & -\mathtt{J}_{\mathrm{s}}\int\frac{\mathrm{d}^{3}k}{(2\pi)^{3}}\,\Big\{\hat{\eta}_{\uparrow}(\mathbf{k})\big[-\cos\theta_{\mathbf{k}}\hat{\varphi}_{\uparrow}^{\dagger}(k_{x})+\sin\theta_{\mathbf{k}}\hat{\varphi}_{\downarrow}(-k_{x})\big]+\hat{\eta}_{\uparrow}^{\dagger}(\mathbf{k})\big[\cos\theta_{\mathbf{k}}\hat{\varphi}_{\uparrow}(k_{x})-\sin\theta_{\mathbf{k}}\hat{\varphi}_{\downarrow}^{\dagger}(-k_{x}\text{)}\big]\\
 & +\hat{\eta}_{\downarrow}(\mathbf{k})\big[-\cos\theta_{\mathbf{k}}\hat{\varphi}_{\downarrow}^{\dagger}(k_{x})-\sin\theta_{\mathbf{k}}\hat{\varphi}_{\uparrow}(-k_{x})\big]+\hat{\eta}_{\downarrow}^{\dagger}(\mathbf{k})\big[\cos\theta_{\mathbf{k}}\hat{\varphi}_{\downarrow}(k_{x})+\sin\theta_{\mathbf{k}}\hat{\varphi}_{\uparrow}^{\dagger}(-k_{x}\text{)}\big]\Big\}.
\end{aligned}
\label{HG}
\end{equation}

In order to obtain the effective theory for the nanowire, we utilize
the Fr\"ohlich-Nakajima transformation to eliminate the quasi-particle
excitation in SC. For the total Hamiltonian $\hat{\mathcal{H}}=[\hat{H}_{\mathrm{w}}+\hat{H}_{\mathrm{sc}}]+\hat{H}_{\mathrm{w\text{-}sc}}:=\hat{H}_{0}+\hat{H}_{1}$,
we apply a unitary transformation
\begin{equation}
\hat{H}_{S}=e^{-\hat{S}}\hat{H}e^{\hat{S}}=\hat{H}_{0}+(\hat{H}_{1}+[\hat{H}_{0},\hat{S}])+\frac{1}{2}[\hat{H}_{1},\hat{S}]+\frac{1}{2}[(\hat{H}_{1}+[\hat{H}_{0},\hat{S}]),\hat{S}]
+\ldots,\label{eq:FN transform}
\end{equation}
and the anti-Hermitian operator $\hat{S}$ should be properly set to make
sure $\hat{H}_{1}+[\hat{H}_{0},\hat{S}]\equiv0$, so the effective
Hamiltonian becomes $\hat{H}_{\mathrm{eff}}=\hat{H}_{0}+\frac{1}{2}[\hat{H}_{1},\hat{S}]$.

Specifically, here we adopt an ansatz that $\hat{S}$ has the following
form
\begin{equation}
\begin{aligned}\hat{S}=\int\frac{\mathrm{d}^{3}k}{(2\pi)^{3}} & \Big\{\hat{\eta}_{\uparrow}(\mathbf{k})\big[A_{\mathbf{k}}\hat{\varphi}_{\uparrow}^{\dagger}(k_{x})+B_{\mathbf{k}}\hat{\varphi}_{\downarrow}(-k_{x})+E_{\mathbf{k}}\hat{\varphi}_{\downarrow}^{\dagger}(k_{x})+F_{\mathbf{k}}\hat{\varphi}_{\uparrow}(-k_{x})\big]\\
 & +\hat{\eta}_{\uparrow}^{\dagger}(\mathbf{k})\big[A_{\mathbf{k}}^{\prime}\hat{\varphi}_{\uparrow}(k_{x})+B_{\mathbf{k}}^{\prime}\hat{\varphi}_{\downarrow}^{\dagger}(-k_{x})+E_{\mathbf{k}}^{\prime}\hat{\varphi}_{\downarrow}(k_{x})+F_{\mathbf{k}}^{\prime}\hat{\varphi}_{\uparrow}^{\dagger}(-k_{x})\big]\\
 & +\hat{\eta}_{\downarrow}(\mathbf{k})\big[C_{\mathbf{k}}\hat{\varphi}_{\downarrow}^{\dagger}(k_{x})+D_{\mathbf{k}}\hat{\varphi}_{\uparrow}(-k_{x})+H_{\mathbf{k}}\hat{\varphi}_{\uparrow}^{\dagger}(k_{x})+L_{\mathbf{k}}\hat{\varphi}_{\downarrow}(-k_{x})\big]\\
 & +\hat{\eta}_{\downarrow}^{\dagger}(\mathbf{k})\big[C_{\mathbf{k}}^{\prime}\hat{\varphi}_{\downarrow}(k_{x})+D_{\mathbf{k}}^{\prime}\hat{\varphi}_{\uparrow}^{\dagger}(-k_{x})+H_{\mathbf{k}}^{\prime}\hat{\varphi}_{\uparrow}(k_{x})+L_{\mathbf{k}}^{\prime}\hat{\varphi}_{\downarrow}^{\dagger}(-k_{x})\big]\Big\}.
\end{aligned}
\label{eq:TransformS}
\end{equation}
Then by using the above Eqs.\,(\ref{eq:kH}, \ref{eq:Ksc}, \ref{HG},
\ref{eq:TransformS}), the condition $\hat{H}_{1}+[\hat{H}_{0},\hat{S}]\equiv0$
gives
\begin{equation}
\begin{aligned} & \int\frac{\mathrm{d}^{3}k}{(2\pi)^{3}}\,\Big\{\big[(-E_{\mathbf{k}}^{\mathrm{sc}}+\epsilon_{k_{x},\downarrow})E_{k}+i\alpha k_{x}A_{k}\big]\hat{\eta}_{\uparrow}(\mathbf{\mathbf{k}})\hat{\varphi}_{\downarrow}^{\dagger}(k_{x})-\big[(E_{\mathbf{k}}^{\mathrm{sc}}-\epsilon_{k_{x},\uparrow})A_{k}+i\alpha k_{x}E_{k}-\mathtt{J}_{\mathrm{s}}\cos\theta_{\mathbf{k}}\big]\hat{\eta}_{\uparrow}(\mathbf{k})\hat{\varphi}_{\uparrow}^{\dagger}(k_{x})\\
 & +\big[(E_{\mathbf{k}}^{\mathrm{sc}}-\epsilon_{k_{x},\uparrow})A_{k}^{\prime}-i\alpha k_{x}E_{\mathbf{k}}^{\prime}-\mathtt{J}_{\mathrm{s}}\cos\theta_{\mathbf{k}}\big]\hat{\eta}_{\uparrow}^{\dagger}(\mathbf{k})\hat{\varphi}_{\uparrow}(k_{x})+\big[(E_{\mathbf{k}}^{\mathrm{sc}}-\epsilon_{k_{x},\downarrow})E_{\mathbf{k}}^{\prime}+i\alpha k_{x}A_{\mathbf{k}}^{\prime}\big]\hat{\eta}_{\uparrow}^{\dagger}(\mathbf{k})\hat{\varphi}_{\downarrow}(k_{x})\\
 & -\big[(E_{\mathbf{k}}^{\mathrm{sc}}+\epsilon_{k_{x},\downarrow})B_{k}+i\alpha k_{x}F_{\mathbf{k}}+\mathtt{J}_{\mathrm{s}}\sin\theta_{\mathbf{k}}\big]\hat{\eta}_{\uparrow}(\mathbf{k})\hat{\varphi}_{\downarrow}(-k_{x})+\big[-(E_{\mathbf{k}}^{\mathrm{sc}}+\epsilon_{k_{x},\uparrow})F_{\mathbf{k}}+i\alpha k_{x}B_{\mathbf{k}}\big]\hat{\eta}_{\uparrow}(\mathbf{k})\hat{\varphi}_{\uparrow}(-k_{x})\\
 & +\big[(E_{\mathbf{k}}^{\mathrm{sc}}+\epsilon_{k_{x},\downarrow})B_{k}^{\prime}-i\alpha k_{x}F_{\mathbf{k}}^{\prime}+\mathtt{J}_{\mathrm{s}}\sin\theta_{\mathbf{k}}\big]\hat{\eta}_{\uparrow}^{\dagger}(\mathbf{k})\hat{\varphi}_{\downarrow}^{\dagger}(-k_{x})+\big[(E_{\mathbf{k}}^{\mathrm{sc}}+\epsilon_{k_{x},\uparrow})F_{\mathbf{k}}^{\prime}+i\alpha k_{x}B_{\mathbf{k}}^{\prime}\big]\hat{\eta}_{\uparrow}^{\dagger}(\mathbf{k})\hat{\varphi}_{\downarrow}(k_{x})\\
 & -\big[(E_{\mathbf{k}}^{\mathrm{sc}}-\epsilon_{k_{x},\downarrow})C_{k}-i\alpha k_{x}H_{\mathbf{k}}-\mathtt{J}_{\mathrm{s}}\cos\theta_{\mathbf{k}}\big]\hat{\eta}_{\downarrow}(\mathbf{k})\hat{\varphi}_{\downarrow}^{\dagger}(k_{x})+\big[-(E_{\mathbf{k}}^{\mathrm{sc}}-\epsilon_{k_{x},\uparrow})H_{\mathbf{k}}-i\alpha k_{x}C_{\mathbf{k}}\big]\hat{\eta}_{\downarrow}(\mathbf{k})\hat{\varphi}_{\downarrow}^{\dagger}(k_{x})\\
 & +\big[(E_{\mathbf{k}}^{\mathrm{sc}}-\epsilon_{k_{x},\downarrow})C_{k}^{\prime}+i\alpha k_{x}H_{\mathbf{k}}^{\prime}-\mathtt{J}_{\mathrm{s}}\sin\theta_{\mathbf{k}}\big]\hat{\eta}_{\downarrow}^{\dagger}(\mathbf{k})\hat{\varphi}_{\downarrow}^{\dagger}(k_{x})+\big[(E_{\mathbf{k}}^{\mathrm{sc}}-\epsilon_{k_{x},\uparrow})H_{\mathbf{k}}^{\prime}-i\alpha k_{x}C_{\mathbf{k}}^{\prime}\big]\hat{\eta}_{\downarrow}^{\dagger}(\mathbf{k})\hat{\varphi}_{\uparrow}(k_{x})\\
 & -\big[(E_{\mathbf{k}}^{\mathrm{sc}}+\epsilon_{k_{x},\uparrow})D_{k}-i\alpha k_{x}L_{\mathbf{k}}-\mathtt{J}_{\mathrm{s}}\sin\theta_{\mathbf{k}}\big]\hat{\eta}_{\downarrow}(\mathbf{k})\hat{\varphi}_{\uparrow}(-k_{x})+\big[-(E_{\mathbf{k}}^{\mathrm{sc}}+\epsilon_{k_{x},\downarrow})L_{\mathbf{k}}-i\alpha k_{x}D_{\mathbf{k}}\big]\hat{\eta}_{\downarrow}(\mathbf{k})\hat{\varphi}_{\downarrow}(k_{x})\\
 & +\big[(E_{\mathbf{k}}^{\mathrm{sc}}+\epsilon_{k_{x},\uparrow})D_{k}^{\prime}+i\alpha k_{x}L_{\mathbf{k}}^{\prime}-\mathtt{J}_{\mathrm{s}}\sin\theta_{\mathbf{k}}\big]\hat{\eta}_{\downarrow}^{\dagger}(\mathbf{k})\hat{\varphi}_{\uparrow}^{\dagger}(-k_{x})+\big[(E_{\mathbf{k}}^{\mathrm{sc}}+\epsilon_{k_{x},\downarrow})L_{\mathbf{k}}^{\prime}-i\alpha k_{x}D_{\mathbf{k}}^{\prime}\big]\hat{\eta}_{\downarrow}^{\dagger}(\mathbf{k})\hat{\varphi}_{\downarrow}^{\dagger}(-k_{x})\Big\}\equiv0,
\end{aligned}
\label{eq:coeffeq}
\end{equation}
with $\epsilon_{k_{x},\uparrow(\downarrow)}\equiv\epsilon_{k_{x}}\pm B$.
The coefficients of each term in the above integral should be zero,
and then the undetermined coefficients in $S$ are solved as
\begin{equation}
\begin{aligned}A_{\mathbf{k}} & =A_{\mathbf{k}}^{\prime}=\frac{\mathtt{J}_{\mathrm{s}}\cos\theta_{\mathbf{k}}[E_{\mathbf{k}}^{\mathrm{sc}}-\epsilon_{k_{x},\downarrow}]}{\Pi_{-}(\mathbf{k})},\quad B_{\mathbf{k}}=B_{\mathbf{k}}^{\prime}=-\frac{\mathtt{J}_{\mathrm{s}}\sin\theta_{\mathbf{k}}[E_{\mathbf{k}}^{\mathrm{sc}}+\epsilon_{k_{x},\uparrow}]}{\Pi_{+}(\mathbf{k})},\\
C_{\mathbf{k}} & =C_{\mathbf{k}}^{\prime}=\frac{\mathtt{J}_{\mathrm{s}}\cos\theta_{\mathbf{k}}[E_{\mathbf{k}}^{\mathrm{sc}}-\epsilon_{k_{x},\uparrow}]}{\Pi_{-}(\mathbf{k})},\quad D_{\mathbf{k}}=D_{\mathbf{k}}^{\prime}=\frac{\mathtt{J}_{\mathrm{s}}\sin\theta_{\mathbf{k}}[E_{\mathbf{k}}^{\mathrm{sc}}+\epsilon_{k_{x},\downarrow}]}{\Pi_{+}(\mathbf{k})},\\
E_{\mathbf{k}} & =-E_{\mathbf{k}}^{\prime}=-H_{\mathbf{k}}=H_{\mathbf{k}}^{\prime}=\frac{i\alpha k_{x}\mathtt{J}_{\mathrm{s}}\cos\theta_{\mathbf{k}}}{\Pi_{-}(\mathbf{k})},\quad F_{\mathbf{k}}=-F_{\mathbf{k}}^{\prime}=L_{\mathbf{k}}=-L_{\mathbf{k}}^{\prime}=-\frac{i\alpha k_{x}\mathtt{J}_{\mathrm{s}}\sin\theta_{\mathbf{k}}}{\Pi_{+}(\mathbf{k})},
\end{aligned}
\label{eq:coefficient}
\end{equation}
with $\Pi_{\pm}(\mathbf{k})=(E_{\mathbf{k}}^{\mathrm{sc}}\pm\epsilon_{k_{x}})^{2}-(B^{2}+\alpha^{2}k_{x}^{2})$.

When the magnetic field is not too strong and the electron tunneling
strength between the nanowire and SC is weak, i.e. $|\Pi_{\pm}(\mathbf{k})|\gg(\mathtt{J}_{\mathrm{s}}\mathrm{k_{F}^{s}})^{2}$
($\mathrm{k_{F}^{s}}$ is Fermi momentum of the SC), the effective
Hamiltonian of the nanowire is further obtained as $\hat{H}_{\mathrm{eff}}=\hat{H}_{0}+\frac{1}{2}[\hat{H}_{1},\hat{S}]$,
which is calculated as follows {[}by using $\{\hat{\eta}_{s}(\mathbf{k}),\hat{\eta}_{s'}^{\dagger}(\mathbf{k}^{\prime})\}_{+}=(2\pi)^{3}\delta_{ss'}\delta(\mathbf{k}-\mathbf{k}^{\prime})$
{]}
\begin{align}
 & \frac{1}{2}[\hat{H}_{1},\hat{S}]=\frac{\mathtt{J}_{\mathrm{s}}}{2}\int\frac{\mathrm{d}^{3}k}{(2\pi)^{3}}\,\bigg\{\big[-\cos\theta_{\mathbf{k}}\hat{\varphi}_{\uparrow}^{\dagger}(k_{x})+\sin\theta_{\mathbf{k}}\hat{\varphi}_{\downarrow}(-k_{x})\big]\big[A_{\mathbf{k}}^{\prime}\hat{\varphi}_{\uparrow}(k_{x}\text{)}+B_{\mathbf{k}}^{\prime}\hat{\varphi}_{\downarrow}^{\dagger}(-k_{x}\text{)}+E_{\mathbf{k}}^{\prime}\varphi_{\downarrow}(k_{x}\text{)}+F_{\mathbf{k}}^{\prime}\varphi_{\uparrow}^{\dagger}(-k_{x}\text{)}\big]\nonumber \\
 & +\big[\cos\theta_{\mathbf{k}}\hat{\varphi}_{\uparrow}(k_{x}\text{)}-\sin\theta_{\mathbf{k}}\hat{\varphi}_{\downarrow}^{\dagger}(-k_{x})\big]\big[A_{\mathbf{k}}\hat{\varphi}_{\uparrow}^{\dagger}(k_{x}\text{)}+B_{\mathbf{k}}\hat{\varphi}_{\downarrow}(-k_{x})+E_{\mathbf{k}}\hat{\varphi}_{\downarrow}^{\dagger}(k_{x}\text{)}+F_{\mathbf{k}}\hat{\varphi}_{\uparrow}(-k_{x}\text{)}\big]\nonumber \\
 & +\big[-\cos\theta_{\mathbf{k}}\hat{\varphi}_{\downarrow}^{\dagger}(k_{x}\text{)}-\sin\theta_{\mathbf{k}}\hat{\varphi}_{\uparrow}(-k_{x})\big]\big[C_{\boldsymbol{k}}^{\prime}\hat{\varphi}_{\downarrow}(k_{x}\text{)}+D_{\mathbf{k}}^{\prime}\hat{\varphi}_{\uparrow}^{\dagger}(-k_{x})+H_{\mathbf{k}}^{\prime}\varphi_{\uparrow}(k_{x}\text{)}+L_{\mathbf{k}}^{\prime}\hat{\varphi}_{\downarrow}^{\dagger}(-k_{x}\text{)}\big]\nonumber \\
 & +\big[\cos\theta_{\mathbf{k}}\hat{\varphi}_{\downarrow}(k_{x})+\sin\theta_{\mathbf{k}}\varphi_{\uparrow}^{\dagger}(-k_{x})\big]\big[C_{\mathbf{k}}\hat{\varphi}_{\downarrow}^{\dagger}(k_{x}\text{)}+D_{\mathbf{k}}\hat{\varphi}_{\uparrow}(-k_{x})+H_{\mathbf{k}}\hat{\varphi}_{\uparrow}^{\dagger}(k_{x}\text{)}+L_{\mathbf{k}}\hat{\varphi}_{\downarrow}(-k_{x})\big]\bigg\}\\
 & =\frac{1}{2}\mathtt{J}_{\mathrm{s}}\int\frac{\mathrm{d}^{3}k}{(2\pi)^{3}}\bigg\{\big[-2A_{\mathbf{k}}\cos\theta_{\mathbf{k}}+2D_{\mathbf{k}}\sin\theta_{\mathbf{k}}\big]\hat{\varphi}_{\uparrow}^{\dagger}(k_{x}\text{)}\hat{\varphi}_{\uparrow}(k_{x}\text{)}+\big[-2B_{\mathbf{k}}\sin\theta_{\mathbf{k}}-2C_{k}\cos\theta_{\mathbf{k}}\big]\hat{\varphi}_{\downarrow}^{\dagger}(k_{x}\text{)}\hat{\varphi}_{\downarrow}(k_{x}\text{)}+\big[2\cos\theta_{\mathbf{k}}E_{\mathbf{k}}\nonumber \\
 & -2\sin\theta_{\mathbf{k}}F_{k}\big]\big[\hat{\varphi}_{\uparrow}^{\dagger}(k_{x})\hat{\varphi}_{\downarrow}(k_{x})-\hat{\varphi}_{\downarrow}^{\dagger}(k_{x})\hat{\varphi}_{\uparrow}(k_{x}\text{)}\big]+\big[A_{\mathbf{k}}\sin\theta_{\mathbf{k}}-B_{\mathbf{k}}\cos\theta_{\mathbf{k}}+C_{\mathbf{k}}\sin\theta_{\mathbf{k}}+D_{\mathbf{k}}\cos\theta_{\mathbf{k}}\big]\big[\hat{\varphi}_{\uparrow}^{\dagger}(k_{x}\text{)}\hat{\varphi}_{\downarrow}^{\dagger}(-k_{x}\text{)}\nonumber \\
 & +\hat{\varphi}_{\downarrow}(-k_{x}\text{)}\hat{\varphi}_{\uparrow}(k_{x}\text{)}\big]+\big[\sin\theta_{\mathbf{k}}E_{\mathbf{k}}+\cos\theta_{\mathbf{k}}F_{\mathbf{k}}\big]\big[\hat{\varphi}_{\uparrow}^{\dagger}(k_{x}\text{)}\hat{\varphi}_{\uparrow}^{\dagger}(-k_{x}\text{)}-\hat{\varphi}_{\uparrow}(-k_{x}\text{)}\hat{\varphi}_{\uparrow}(k_{x}\text{)}+\hat{\varphi}_{\downarrow}^{\dagger}(k_{x}\text{)}\hat{\varphi}_{\downarrow}^{\dagger}(-k_{x}\text{)}-\hat{\varphi}_{\downarrow}(-k_{x}\text{)}\hat{\varphi}_{\downarrow}(k_{x}\text{)}\big]\bigg\}.\nonumber
\end{align}
Finally, by substituting the coefficients obtained in Eq.\,(\ref{eq:coefficient})
here, the effective Hamiltonian is obtained
\begin{equation}
\begin{aligned}\hat{H}_{\mathrm{eff}}=\int\frac{\mathrm{d}k}{2\pi} & \,\bigg\{\tilde{\epsilon}_{k}\big[\hat{\varphi}_{\uparrow}^{\dagger}(k_{x})\hat{\varphi}_{\uparrow}(k_{x})+\hat{\varphi}_{\downarrow}^{\dagger}(k_{x})\hat{\varphi}_{\downarrow}(k_{x})\big]+i\tilde{\alpha}k_{x}\big[\hat{\varphi}_{\downarrow}^{\dagger}(k_{x})\hat{\varphi}_{\uparrow}(k_{x})-\mathrm{H.c.}\big]\\
 & +\tilde{B}_{k_{x}}\big[\hat{\varphi}_{\uparrow}^{\dagger}(k_{x})\hat{\varphi}_{\uparrow}(k_{x})-\hat{\varphi}_{\downarrow}^{\dagger}(k_{x})\hat{\varphi}_{\downarrow}(k_{x})\big]+\tilde{\Delta}_{k_{x}}\big[\hat{\varphi}_{\uparrow}^{\dagger}(k_{x})\hat{\varphi}_{\downarrow}^{\dagger}(-k_{x})+\mathrm{H.c.}\big]\\
 & +\tilde{\Lambda}_{k_{x}}\big[\hat{\varphi}_{\downarrow}^{\dagger}(k_{x})\hat{\varphi}_{\downarrow}^{\dagger}(-k_{x})+\hat{\varphi}_{\uparrow}^{\dagger}(k_{x})\hat{\varphi}_{\uparrow}^{\dagger}(-k_{x})-\mathrm{H.c.}\big]\bigg\}.
\end{aligned}
\end{equation}
The electron tunneling between the nanowire and the \emph{s}-wave
SC modifies the kinetic energy $\tilde{\epsilon}_{k_{x}}$, Zeeman
splitting $\tilde{B}_{k_{x}}$ and spin-orbit coupling $\tilde{\alpha}_{k_{x}}$,
and these parameters are
\begin{equation}
\begin{aligned}\tilde{B}_{k_{x}} & =[1-\mathtt{J}_{\mathrm{s}}^{2}\int\frac{\mathrm{d}k_{y}\mathrm{d}k_{z}}{(2\pi)^{2}}(\frac{\cos^{2}\theta_{\mathbf{k}}}{\Pi_{-}(\mathbf{k})}+\frac{\sin^{2}\theta_{\mathbf{k}}}{\Pi_{+}(\mathbf{k})})]B,\\
\tilde{\alpha}_{k_{x}} & =[1-\mathtt{J}_{\mathrm{s}}^{2}\int\frac{\mathrm{d}k_{y}\mathrm{d}k_{z}}{(2\pi)^{2}}(\frac{\cos^{2}\theta_{\mathbf{k}}}{\Pi_{-}(\mathbf{k})}+\frac{\sin^{2}\theta_{\mathbf{k}}}{\Pi_{+}(\mathbf{k})})]\alpha,\\
\tilde{\epsilon}_{k_{x}} & =\epsilon_{k_{x}}-\mathtt{J}_{\mathrm{s}}^{2}\int\frac{\mathrm{d}k_{y}\mathrm{d}k_{z}}{(2\pi)^{2}}[\frac{\cos^{2}\theta_{\mathbf{k}}E_{-}^{\mathrm{s}}(\mathbf{k})}{\Pi_{-}(\mathbf{k})}-\frac{\sin^{2}\theta_{\mathbf{k}}E_{\text{+}}^{\mathrm{s}}(\mathbf{k})}{\Pi_{+}(\mathbf{k})}]\\
\tilde{\Delta}_{k_{x}} & =\mathtt{J}_{\mathrm{s}}^{2}\int\frac{\mathrm{d}k_{y}\mathrm{d}k_{z}}{(2\pi)^{2}}\frac{\sin2\theta_{\mathbf{k}}}{2}[\frac{E_{-}^{\mathrm{s}}(\mathbf{k})}{\Pi_{-}(\mathbf{k})}+\frac{E_{\text{+}}^{\mathrm{s}}(\mathbf{k})}{\Pi_{+}(\mathbf{k})}],\\
\tilde{\Lambda}_{k_{x}} & =\mathtt{J}_{\mathrm{s}}^{2}\int\frac{\mathrm{d}k_{y}\mathrm{d}k_{z}}{(2\pi)^{2}}\frac{i\alpha k_{x}\sin2\theta_{\mathbf{k}}}{4}(\frac{1}{\Pi_{-}(\mathbf{k})}-\frac{1}{\Pi_{+}(\mathbf{k})}),
\end{aligned}
,\label{eq:corrected coeffecint}
\end{equation}
where $E_{\text{\ensuremath{\pm}}}^{\mathrm{s}}(\mathbf{k})\equiv E_{\mathbf{k}}^{\mathrm{sc}}\pm\epsilon_{k_{x}}$,
and $\Pi_{\pm}(\mathbf{k})$, $\sin\theta_{\mathbf{k}}$, $\cos\theta_{\mathbf{k}}$
have been given in Eqs.\,(\ref{eq:bogoliubov transform}, \ref{eq:coefficient})
respectively.

To further calculate the integrals in the above result, we consider
that the Zeeman splitting \textbf{$B$} and spin-orbital coupling
$\alpha$ of the nanowire are much smaller than the \emph{s}-wave
SC gap, thus $|E_{\mathbf{k}}^{\mathrm{sc}}-\sqrt{B^{2}+\alpha^{2}k_{x}^{2}}|\gg|\epsilon_{k_{x}}|$.
Then in the above integrals we omit the second and higher-order terms
of $|\alpha k_{x}|/[E_{\mathbf{k}}^{\mathrm{sc}}+\sqrt{B^{2}+\alpha^{2}k_{x}^{2}}]$
and $|\epsilon_{k_{x}}|/[E_{\mathbf{k}}^{\mathrm{sc}}-\sqrt{B^{2}+\alpha^{2}k_{x}^{2}}]$,
and the above parameters in Eq.\,(\ref{eq:corrected coeffecint})
are simplified as
\begin{align}
\tilde{\Delta}_{k_{x}} & =\Upsilon_{\mathrm{s}}(1-\frac{\alpha^{2}k_{x}^{2}+B^{2}}{\Delta_{\mathrm{s}}^{2}})^{-\frac{1}{2}},\quad\tilde{\Lambda}_{k_{x}}=0\label{eq:modifed coeff}\\
\frac{\tilde{B}_{k_{x}}}{B} & =\frac{\tilde{\alpha}_{k_{x}}}{\alpha}=\frac{\tilde{\epsilon}_{k_{x}}}{\epsilon_{k_{x}}}=\frac{\tilde{\mu}_{k_{x}}}{\mu}=1-\frac{\tilde{\Delta}_{k_{x}}}{\Delta_{\mathrm{s}}}.\nonumber
\end{align}
Here, $\Upsilon_{\mathrm{s}}=\mathtt{J}_{\mathrm{s}}^{2}\rho_{\mathrm{s}}$
describes the coupling strength between the \emph{s}-wave SC and the
nanowire, where $\rho_{\mathrm{s}}$ is two-dimensional superconducting
density of states. Now we have obtained the effective Hamiltonian
of the nanowire in the main text, i.e.,
\begin{equation}
\begin{aligned}\hat{H}_{\mathrm{eff}}=\int\frac{\mathrm{d}k_{x}}{2\pi} & \,\bigg\{\tilde{\epsilon}_{k_{x}}\big[\hat{\varphi}_{\uparrow}^{\dagger}(k_{x})\hat{\varphi}_{\uparrow}(k_{x})+\hat{\varphi}_{\downarrow}^{\dagger}(k_{x})\hat{\varphi}_{\downarrow}(k_{x})\big]+i\tilde{\alpha}_{k_{x}}k_{x}\big[\hat{\varphi}_{\downarrow}^{\dagger}(k_{x})\hat{\varphi}_{\uparrow}(k_{x})-\mathrm{H.c.}\big]\\
 & +\tilde{B}_{k_{x}}\big[\hat{\varphi}_{\uparrow}^{\dagger}(k_{x})\hat{\varphi}_{\uparrow}(k_{x})-\hat{\varphi}_{\downarrow}^{\dagger}(k_{x})\hat{\varphi}_{\downarrow}(k_{x})\big]+\tilde{\Delta}_{k_{x}}\big[\hat{\varphi}_{\uparrow}^{\dagger}(k_{x})\hat{\varphi}_{\downarrow}^{\dagger}(-k_{x})+\mathrm{H.c.}\big]\bigg\}.
\end{aligned}
\label{eq:KHeff}
\end{equation}
Clearly, $\tilde{\Delta}_{k_{x}}$ is the pairing potential induced
by the SC proximity effect, and all these parameters $\tilde{\epsilon}_{k_{x}}$,
$\tilde{\alpha}_{k_{x}}$, $\tilde{B}_{k_{x}}$, $\tilde{\Delta}_{k_{x}}$
exhibit significant dependence on the magnetic field. In addition,
when the SC gap is so large that $|E_{\mathbf{k}}^{\mathrm{sc}}|\gg B,\,\Upsilon_{\mathrm{s}}$,
in the low energy regime $(k\simeq0)$, the induced pairing strength
can be approximated as a constant $\tilde{\Delta}\simeq\Upsilon_{\mathrm{s}}$,
and the kinetic energy, spin-orbital coupling and Zeeman splitting
of the nanowire are corrected by the constant factor $1-\Upsilon_{\mathrm{s}}/\Delta_{\mathrm{s}}$
according to (\ref{eq:modifed coeff}).

Then the quasi-particle energy spectrum determined by the effective
Hamiltonian in Eq.\,(\ref{eq:KHeff}) is
\begin{equation}
\begin{aligned}E_{\pm}(k_{x})=\sqrt{\tilde{\Delta}_{k_{x}}^{2}+\frac{\tilde{\epsilon}_{k_{x},+}^{2}+\tilde{\epsilon}_{k_{x},-}^{2}}{2}\pm(\tilde{\epsilon}_{k_{x},+}-\tilde{\epsilon}_{k_{x},-})\sqrt{[\tilde{\Delta}_{k_{x}}^{(\mathrm{s})}]^{2}+\tilde{\epsilon}_{k_{x}}^{2}}},\end{aligned}
\end{equation}
with $\tilde{\epsilon}_{k_{x},\pm}=\tilde{\epsilon}_{k_{x}}\pm\sqrt{\tilde{B}_{k_{x}}^{2}+\tilde{\alpha}_{k_{x}}^{2}k_{x}^{2}}$
and $\tilde{\Delta}_{k_{x}}^{(\mathrm{s})}=B\tilde{\Delta_{k_{x}}}/\sqrt{B^{2}+\alpha^{2}k_{x}^{2}}$.
The energy level crossing point of the quasi-particle energy spectrum
is $E_{-}(k_{x}=0)=0$, which gives the critical condition of topological
phase: $\tilde{B}_{k_{x}=0}=\sqrt{\tilde{\mu}_{k_{x}=0}^{2}+\tilde{\Delta}_{k_{x}=0}^{2}}$.
When the corrected magnetic field satisfies
\begin{equation}
\begin{aligned}\tilde{B}_{k_{x}=0}(B,\Upsilon_{\mathrm{s}})>\sqrt{\tilde{\mu}_{k_{x}=0}^{2}(B,\Upsilon_{\mathrm{s}})+\tilde{\Delta}_{k_{x}=0}^{2}(B,\Upsilon_{\mathrm{s}})},\end{aligned}
\label{eq:topological condiction}
\end{equation}
the energy gap reopens, and the topological phase with Majorana zero
modes emerges in the region. However, when the magnetic field gets
stronger, the corrected magnetic field decreases significantly according
to Eq. (\ref{eq:modifed coeff}), which makes the corrected magnetic
field become $\tilde{B}_{k_{x}=0}<\sqrt{\tilde{\mu}_{k_{x}=0}^{2}+\tilde{\Delta}_{k_{x}=0}^{2}}$.
Then the topological phase may disappear. Therefore, the topological
phase only emerge at proper magnetic field strength.

\section{Majorana zero modes determined by the effective Hamiltonian of the
nanowire}

Here, based on the effective Hamiltonian of the nanowire, we give
the region of the existence of Majorana zero modes in the $\mu-B$
diagram, which is consistent with the topological phase region given
by Eq.\,(\ref{eq:topological condiction}). In the continuous limit,
the effective Hamiltonian (\ref{eq:KHeff}) of the nanowire in the
low energy regime $(k_{x}\simeq0)$ becomes
\begin{equation}
\hat{H}_{\mathrm{eff}}=\int\mathrm{d}x\,\hat{\boldsymbol{\psi}}^{\dagger}(x)(1-Z(B))[-\frac{\partial_{x}^{2}}{2m_{\mathrm{w}}}-\mu-i\alpha\sigma^{y}\partial_{x}+B\sigma^{z}]\hat{\boldsymbol{\psi}}(x)+Z(B)\Upsilon_{\mathrm{s}}[\hat{\psi}_{\uparrow}^{\dagger}(x)\hat{\psi}_{\downarrow}^{\dagger}(x)+\hat{\psi}_{\downarrow}(x)\hat{\psi}_{\uparrow}(x)],\label{eq:nanowire Heff}
\end{equation}
where the corrected factor is defined as $Z(B)\equiv(1-\frac{B^{2}}{\Delta_{\mathrm{s}}^{2}})^{-\frac{1}{2}}$.
In the Nambu representation, the above Hamiltonian becomes
\begin{equation}
\hat{H}_{\mathrm{eff}}=\frac{1}{2}\int\mathrm{d}x\,\hat{\mathbf{\Phi}}^{\dagger}(x)\cdot\mathbf{H}_{x}\cdot\hat{\mathbf{\Phi}}(x),\label{eq:H in Nanbu}
\end{equation}
with $\hat{\mathbf{\Phi}}(x)=[\hat{\boldsymbol{\psi}}(x),\hat{\boldsymbol{\psi}}^{\dagger}(x)]^{T}$
as the 4-component operator and
\begin{equation}
\mathbf{H}_{x}=\begin{bmatrix}[1-Z(B)](-\frac{\partial_{x}^{2}}{2m_{\mathrm{w}}}-\mu-i\alpha\sigma^{y}\partial_{x}+B\sigma^{z}) & i\sigma^{y}Z(B)\Upsilon_{\mathrm{s}}\\
-i\sigma^{y}Z(B)\Upsilon_{\mathrm{s}} & [1-Z(B)](\frac{\partial_{x}^{2}}{2m_{\mathrm{w}}}+\mu+i\alpha\sigma^{y}\partial_{x}-B\sigma^{z})
\end{bmatrix}.\label{eq:single particleH}
\end{equation}
Considering the Bogoliubov-de-Gennes (BdG) equation
\begin{equation}
\mathbf{H}_{x}\boldsymbol{\Psi}_{E}(x)=E\boldsymbol{\Psi}_{E}(x),\label{eq:eigeneq}
\end{equation}
the corresponding wave function is $\boldsymbol{\Psi}_{E}(x)=[u_{\uparrow,E}(x),u_{\downarrow,E}(x),v_{\uparrow,E}(x),v_{\downarrow,E}(x)]^{T}.$
Then the effective Hamiltonian (\ref{eq:nanowire Heff}) is diagonalized
as $H_{\mathrm{eff}}=\frac{1}{2}\sum_{E}E\,\hat{\gamma}_{E}^{\dagger}\hat{\gamma}_{E}$,
where
\begin{equation}
\hat{\gamma}_{E}=\int\mathrm{d}x\,\text{\ensuremath{\sum_{s}[u_{s,E}^{*}(x)\hat{\psi_{s}}(x)+v_{s,E}^{*}(x)\hat{\psi}_{s}^{\dagger}(x)].}}\label{eq:quasiparticle}
\end{equation}

For the Majorana fermion, the antiparticle is itself, which means
the quasi-particle operator is self-Hermitian $\hat{\gamma}_{E}=\hat{\gamma}_{E}^{\dagger}$, namely,
\begin{equation}
u_{s,E}(x)=v_{s,E}^{*}(x).\label{eq:uvrelation1}
\end{equation}
 And due to the particle-hole symmetry of the system, the single-particle
Hamiltonian satisfies $\sigma^{x}\mathbf{H}_{x}^{T}\sigma^{x}=-\mathbf{H}_{x}$.
So $\sigma^{x}\boldsymbol{\Psi}_{E}^{*}(x)$ is the eigenstate of
the single-particle Hamiltonian for the energy $-E$:
\begin{equation}
\mathbf{H}_{x}[\sigma^{x}\boldsymbol{\Psi}_{E}^{*}(x)]=-E[\sigma^{x}\boldsymbol{\Psi}_{E}^{*}(x)].\label{eq:-Egien eq}
\end{equation}
Correspondingly, the quasi-particle operator is
\begin{equation}
\hat{\gamma}_{-E}=\int\mathrm{d}x\,\sum_{s}[v_{s,E}\hat{\psi}_{s}(x)+u_{s,E}(x)\hat{\psi}_{s}^{\dagger}(x)].\label{eq:quasiparticle-E}
\end{equation}
If the wave function $\boldsymbol{\Psi}_{E}(x)$ is non-degenerate,
we obtain relation of $u_{s,E}(x)$ and $v_{s,E}(x)$ by Eqs.\,(\ref{eq:quasiparticle},
\ref{eq:quasiparticle-E}) again, i.e.,
\begin{equation}
u_{s,E}(x)=v_{s,-E}^{*}(x).\label{eq:uvrelation2}
\end{equation}
Due to the self-Hermitian relation of the Majorana fermion (\ref{eq:uvrelation1})
and the particle-hole symmetry of the system (\ref{eq:uvrelation2}),
the components of the wave function $u_{s,E}(x)$ and $v_{s,E}(x)$
satisfy
\begin{equation}
u_{s,E}(x)=u_{s,-E}(x).\label{eq:uE cond}
\end{equation}
Thus, only the zero mode quasi-particle could be self-Hermitian, i.e.
the Majorana zero-mode.

For $E=0$, considering that the Majorana fermion requires $u_{s,E}(x)=v_{s,E}^{*}(x)$,
the BdG equation (\ref{eq:eigeneq}) is reduced to
\begin{equation}
[1-Z(B)](-\frac{\partial_{x}^{2}}{2m_{\mathrm{w}}}-\mu-i\alpha\sigma^{y}\partial_{x}+B\sigma^{z})\boldsymbol{u}(x)+i\sigma^{y}Z(B)\Upsilon_{\mathrm{s}}\boldsymbol{u}^{*}(x)=0,\label{eq:two component eq}
\end{equation}
with the 2-component wave function $\boldsymbol{u}(x)\equiv[u_{\uparrow}(x),u_{\downarrow}(x)]^{T}$.
Here we consider the length of the nanowire is $L$, and $x\in[0,L]$.
Then $\boldsymbol{u}(x)$ can be decomposed into the real and imaginary
parts $\boldsymbol{u}(x)=\boldsymbol{u}^{(r)}(x)+i\boldsymbol{u}^{(i)}(x)$,
and we assume they have the following forms
\begin{equation}
\begin{aligned}\boldsymbol{u}^{(r)}(x) & =e^{-\xi_{r}x}[u_{\uparrow}^{(r)},u_{\downarrow}^{(r)}]^{T},\\
\boldsymbol{u}^{(i)}(x) & =e^{-\xi_{i}x}[u_{\uparrow}^{(i)},u_{\downarrow}^{(i)}]^{T},
\end{aligned}
\label{eq:solution}
\end{equation}
where $\xi_{r}$ and $\xi_{i}$ are real numbers. Taking the ansatz
(\ref{eq:solution}) into the reduced BdG equation (\ref{eq:two component eq}),
we get the equations of the real and imaginary parts respectively
\begin{equation}
\begin{aligned}\begin{bmatrix}-\frac{\xi_{r}^{2}}{2m_{\mathrm{w}}}-\mu+B & (-\alpha\xi_{r}+\frac{Z(B)}{1-Z(B)}\Upsilon_{\mathrm{s}})\\
(\alpha\xi_{r}-\frac{Z(B)}{1-Z(B)}\Upsilon_{\mathrm{s}}) & -\frac{\xi_{r}^{2}}{2m_{\mathrm{w}}}-\mu-B
\end{bmatrix}\boldsymbol{u}^{(r)}(x) & =0,\\
\begin{bmatrix}-\frac{\xi_{i}^{2}}{2m_{\mathrm{w}}}-\mu+B & (-\alpha\xi_{i}-\frac{Z(B)}{1-Z(B)}\Upsilon_{\mathrm{s}})\\
(\alpha\xi_{i}+\frac{Z(B)}{1-Z(B)}\Upsilon_{\mathrm{s}}) & -\frac{\xi_{i}^{2}}{2m_{\mathrm{w}}}-\mu-B
\end{bmatrix}\boldsymbol{u}^{(i)}(x) & =0.
\end{aligned}
\label{eq:urieq}
\end{equation}
If the above two equations have solutions, the determinants of the
coefficient matrices in Eq.\,(\ref{eq:urieq}) are zero
\begin{align}
0=[-\frac{1}{2m_{\mathrm{w}}}\xi_{r}^{2}-\mu]^{2}+(\alpha\xi_{r}-\frac{Z(B)}{1-Z(B)}\Upsilon_{\mathrm{s}})^{2}-B^{2} & \equiv f(\xi_{r}),\label{eq:Beq}\\
0=[-\frac{1}{2m_{\mathrm{w}}}\xi_{i}^{2}-\mu]^{2}+(\alpha\xi_{i}+\frac{Z(B)}{1-Z(B)}\Upsilon_{\mathrm{s}})^{2}-B^{2} & \equiv g(\xi_{i}).\label{eq:B2eq}
\end{align}
Here, the parameters $\mu$, $m_{\mathrm{w}}$, $\alpha$ and $B$
are all positive, and the corrected factor satisfies $0<Z(B)<1$,
and thus $Z(B)/[1-Z(B)]>0$.

(a) If the equation $f(\xi)=0$ has a positive root $\xi_{r}>0$,
the real part $\boldsymbol{u}^{(r)}(x)$ has a solution localized
around the end $x=0$. The existence condition for $\xi_{r}$ can
be given by examining the monotonicity $f(\xi)$. Notice that the
quartic function $f(\xi)$ satisfies
\begin{equation}
f^{\prime\prime}(\xi)=\frac{1}{m}[\frac{3}{m}\xi^{2}+2\mu]+2\alpha^{2}>0,\label{eq:function f}
\end{equation}
thus $f^{\prime}(\xi)$ is monotonically increasing. Thus, in the
interval $\xi\in[0,\infty)$, the minimum of $f^{\prime}(\xi)$ appears
at $\xi=0$, which is $f^{\prime}(0)=-2\alpha Z(B)\Upsilon_{\mathrm{s}}/[1-Z(B)]<0$.
Therefore, there must exist a certain $\xi_{0}>0$ satisfying $f'(\xi_{0})=0$,
that means, when $0\le\xi<\xi_{0}$, $f(\xi)$ is monotonically decreasing,
and when $\xi>\xi_{0}$, $f(\xi)$ is monotonically increasing. Namely,
the minimum of $f(\xi)$ appears at $\xi_{0}$. To make sure Eq.\,(\ref{eq:Beq})
have one solution or two solutions, we must have $\min f(\xi)=f(\xi_{0})<0$.

(b) If the equation $g(\xi)=0$ has a positive root $\xi_{i}>0$,
the imaginary part $\boldsymbol{u}^{(i)}(x)$ has a solution localized
around the end $x=0$. In the interval $\xi\in[0,\infty)$, $g^{\prime}(\xi)$
is always positive, i.e.,
\begin{equation}
g^{\prime}(\xi)=\frac{2\xi}{m}[\frac{1}{2m}\xi^{2}+\mu]+2\alpha(\alpha\xi+\frac{Z(B)}{1-Z(B)}\Upsilon_{\mathrm{s}})>0,\label{eq:derivative}
\end{equation}
thus $g(\xi)$ is monotonically increasing. Therefore, to make sure
Eq.\,(\ref{eq:B2eq}) have a solution, we must have $\min g(\xi)=g(0)<0$.

Now combining the two existence conditions from (a) and (b), since
$\min g(\xi)=g(0)=f(0)>\min f(\xi)$, to make sure both $\boldsymbol{u}^{(r)}(x)$
and $\boldsymbol{u}^{(i)}(x)$ have solutions, we must have $\min g(\xi)=g(0)<0$,
and that gives
\begin{equation}
B>\sqrt{\mu^{2}+[\frac{Z(B)}{1-Z(B)}\Upsilon_{\mathrm{s}}]^{2}}.\label{eq:condition}
\end{equation}

Similarly, it is also proved that there is an edge state around the
other end ($x=L$) if Eq.\,(\ref{eq:condition}) is satisfied, which
means the existence of the Majorana zero modes localized at the two
ends. Notice that this condition is just the same with the topological
phase criterion (\ref{eq:topological condiction}) obtained from crossing
point of the quasi-particle energy spectrum.

\section{Hamiltonian description for the nanowire system}

Now we study the transport behavior of the nanowire system. We consider
the nanowire is contacted with two electron leads at the two ends,
and then derive a quantum Langevin equation to describe the dynamics
of the nanowire. First, the Hamiltonian (\ref{eq:nonawireH}) of the
nanowire is discretized as $N$ sites, that is
\begin{align}
\hat{H}_{\mathrm{w}}= & \sum_{n,s}-\frac{J}{2}(\hat{d}_{n,s}^{\dagger}\hat{d}_{n+1,s}+\hat{d}_{n+1,s}^{\dagger}\hat{d}_{n,s})-(\mu-J)\hat{d}_{n,s}^{\dagger}\hat{d}_{n,s}\nonumber \\
 & +\sum_{n}\frac{\alpha}{2}(\hat{d}_{n,\downarrow}^{\dagger}\hat{d}_{n+1,\uparrow}-\hat{d}_{n,\uparrow}^{\dagger}\hat{d}_{n+1,\downarrow}+\mathrm{H.c.})+\sum_{n}B(\hat{d}_{n,\uparrow}^{\dagger}\hat{d}_{n,\uparrow}-\hat{d}_{n,\downarrow}^{\dagger}\hat{d}_{n,\downarrow}).\label{eq:H-w-1}
\end{align}
Here, $s=\uparrow,\downarrow$ indexes the electron spin, $\alpha$
indicates the spin-orbit coupling strength, $J$ the hopping amplitude,
$\mu$ the chemical potential of the nanowire, and $B$ the Zeeman
splitting from the external magnetic field.

All the $N$ sites are contacted with an \emph{s}-wave SC independently,
and two electron leads are contacted with site-$1$ and site-$N$.
The \emph{s}-wave SC and the two leads are treated as the fermion
baths of the nanowire, and they are described by
\begin{align}
\hat{H}_{\mathrm{sc}} & =\sum_{\mathbf{k}}\epsilon_{\mathbf{k}}^{\mathrm{sc}}(\hat{c}_{\mathbf{k}\uparrow}^{\dagger}\hat{c}_{\mathbf{k}\uparrow}-\hat{c}_{-\mathbf{k}\downarrow}\hat{c}_{-\mathbf{k}\downarrow}^{\dagger})+\Delta_{\mathrm{s}}(\hat{c}_{\mathbf{k}\uparrow}^{\dagger}\hat{c}_{-\mathbf{k}\downarrow}^{\dagger}+\mathrm{H.c.}),\nonumber \\
\hat{H}_{\text{e-}x} & =\sum_{\mathbf{k}s}\varepsilon_{x,\mathbf{k}}\hat{b}_{x,\mathbf{k}s}^{\dagger}\hat{b}_{x,\mathbf{k}s},\qquad(x=1,N)\label{eq:H-sc,e}
\end{align}
Both the two leads and the \emph{s}-wave SC are coupled with the nanowire
through the tunneling interaction {[}see also Eq.\,(\ref{eq:tunneling}){]},
\begin{equation}
\hat{H}_{\text{w-sc}}=-\sum_{n,\mathbf{k}s}(\mathtt{J}_{n,\mathbf{k}}\,\hat{d}_{n,s}^{\dagger}\hat{c}_{\mathbf{k}s}+\mathrm{H.c.}),\qquad\hat{H}_{\text{w-}x}=-\sum_{\mathbf{k}s}(\mathtt{g}_{x,\mathbf{k}}\,\hat{d}_{x,s}^{\dagger}\hat{b}_{x,\mathbf{k}s}+\mathrm{H.c.}).
\end{equation}
Denoting $[\hat{\mathbf{d}}(t)]_{1\times4N}:=(\hat{\boldsymbol{d}}_{1},\dots,\hat{\boldsymbol{d}}_{N})^{T}$
with blocks $[\hat{\boldsymbol{d}}_{n}]_{1\times4}:=(\hat{d}_{n\uparrow},\hat{d}_{n\downarrow},\hat{d}_{n\uparrow}^{\dagger},\hat{d}_{n\downarrow}^{\dagger})^{T}$,
$\hat{\mathbf{c}}_{\mathbf{k}}(t):=(\hat{c}_{\mathbf{k}\uparrow},\hat{c}_{-\mathbf{k}\downarrow},\hat{c}_{\mathbf{k}\uparrow}^{\dagger},\hat{c}_{-\mathbf{k}\downarrow}^{\dagger})^{T}$,
and $\hat{\mathbf{b}}_{x,\mathbf{k}}(t):=(\hat{b}_{x,\mathbf{k}\uparrow},\hat{b}_{x,\mathbf{k}\downarrow},\hat{b}_{x,\mathbf{k}\uparrow}^{\dagger},\hat{b}_{x,\mathbf{k}\downarrow}^{\dagger})^{T}$,
the above Hamiltonians (\ref{eq:H-w-1}, \ref{eq:H-sc,e}) can be
rewritten as
\begin{gather}
\hat{H}_{\mathrm{w}}=\frac{1}{2}\hat{\mathbf{d}}^{\dagger}\cdot\mathbf{H}_{\mathrm{w}}\cdot\hat{\mathbf{d}},\qquad\hat{H}_{\mathrm{sc}}=\frac{1}{2}\sum_{\mathbf{k}}\hat{\mathbf{c}}_{\mathbf{k}}^{\dagger}\cdot\mathbf{H}_{\mathbf{k}}^{\mathrm{sc}}\cdot\hat{\mathbf{c}}_{\mathbf{k}},\qquad\hat{H}_{\text{e-}x}=\frac{1}{2}\hat{\mathbf{b}}_{x,\mathbf{k}}^{\dagger}\cdot\mathbf{H}_{\mathbf{k}}^{\text{e-}x}\cdot\hat{\mathbf{b}}_{x,\mathbf{k}},\nonumber \\
\mathbf{H}_{\mathbf{k}}^{\mathrm{sc}}:=\left[\begin{array}{cccc}
\epsilon_{\mathbf{k}}^{\mathrm{sc}} &  &  & \Delta_{\mathrm{s}}\\
 & \epsilon_{\mathbf{k}}^{\mathrm{sc}} & -\Delta_{\mathrm{s}}\\
 & -\Delta_{\mathrm{s}} & -\epsilon_{\mathbf{k}}^{\mathrm{sc}}\\
\Delta_{\mathrm{s}} &  &  & -\epsilon_{\mathbf{k}}^{\mathrm{sc}}
\end{array}\right],\qquad\mathbf{H}_{\mathbf{k}}^{\text{e-}x}:=\left[\begin{array}{cccc}
\varepsilon_{x,\mathbf{k}}\\
 & \varepsilon_{x,\mathbf{k}}\\
 &  & -\varepsilon_{x,\mathbf{k}}\\
 &  &  & -\varepsilon_{x,\mathbf{k}}
\end{array}\right].
\end{gather}

The system dynamics can be given by the Heisenberg equation. Here,
we consider all the dynamical observables are corrected as $\hat{o}(t)=\hat{o}(t)\Theta(t)e^{-\epsilon t}:=\hat{o}(t)\Theta^{(\epsilon)}(t)$
with $\epsilon\rightarrow0$, namely, the dynamical evolution starts
from $t=0$. The infinitesimal $\epsilon$ guarantees the convergence
of the evolution, and would naturally lead to the causality in the
dynamical propagator. Then the equations of motions becomes $\partial_{t}[\hat{o}(t)\Theta^{(\epsilon)}(t)]=\delta(t)\hat{o}(0)-i\Theta^{(\epsilon)}(t)[\hat{o},\hat{\mathcal{H}}]$,
with $\hat{o}(0)$ as the initial state. Then the dynamics for the
nanowire $\hat{d}_{n,s}$, \emph{s}-wave SC $\hat{c}_{\mathbf{k}s}$,
and electron leads $\hat{b}_{x,\mathbf{k}s}$ are given as
\begin{align}
\partial_{t}\hat{d}_{n,s} & =\delta(t)\hat{d}_{n,s}(0)-i[\hat{d}_{n,s},\hat{H}_{\mathrm{w}}]+i\sum_{\mathbf{k}}\mathtt{J}_{n,\mathbf{k}}\hat{c}_{\mathbf{k},s}+i\sum_{x=1,N}\sum_{\mathbf{k}}\mathtt{g}_{x,\mathbf{k}}\hat{b}_{x,\mathbf{k}s},\nonumber \\
\partial_{t}\hat{c}_{\mathbf{k},s} & =\delta(t)\hat{c}_{\mathbf{k},s}(0)-i[\hat{c}_{\mathbf{k},s},\hat{H}_{\mathrm{sc}}]+i\sum_{n}\mathtt{J}_{n,\mathbf{k}}^{*}\hat{d}_{n,s},\label{eq:Heisenberg}\\
\partial_{t}\hat{b}_{x,\mathbf{k}s} & =\delta(t)\hat{b}_{x,\mathbf{k}s}(0)-i\varepsilon_{x,\mathbf{k}}\hat{b}_{x,\mathbf{k}s}+i\mathtt{g}_{x,\mathbf{k}}^{*}\hat{d}_{x,s}.\nonumber
\end{align}

These dynamical equations can be solved in the Fourier space. We adopt
the following Fourier transform,
\begin{alignat}{1}
\hat{d}(t) & =\int_{-\infty}^{\infty}\frac{\mathrm{d}\omega}{2\pi}\,\tilde{d}(\omega)e^{-i\omega t},\qquad\tilde{d}(\omega)=\int_{-\infty}^{\infty}\mathrm{d}t\,\hat{d}(t)e^{+i\omega t},\nonumber \\
\hat{d}^{\dagger}(t) & =\int_{-\infty}^{\infty}\frac{\mathrm{d}\omega}{2\pi}\,[\tilde{d}(\omega)]^{\dagger}e^{+i\omega t}=\int_{-\infty}^{\infty}\frac{\mathrm{d}\bar{\omega}}{2\pi}\,\tilde{d}^{\dagger}(-\bar{\omega})e^{-i\bar{\omega}t}.
\end{alignat}
Under this convention, the Fourier images for the vectors $\hat{\boldsymbol{d}}_{n}(t)$,
$\hat{\mathbf{c}}_{\mathbf{k}}(t)$, $\hat{\mathbf{b}}_{x,\mathbf{k}}(t)$
read
\begin{align}
\tilde{\mathbf{d}}_{n}(\omega) & =\big(\tilde{d}_{n\uparrow}(\omega),\,\tilde{d}_{n\downarrow}(\omega),\,\tilde{d}_{n\uparrow}^{\dagger}(-\omega),\,\tilde{d}_{n\downarrow}^{\dagger}(-\omega)\big)^{T},\nonumber \\
\tilde{\mathbf{c}}_{\mathbf{k}}(\omega) & =\big(\tilde{c}_{\mathbf{k}\uparrow}(\omega),\,\tilde{c}_{-\mathbf{k}\downarrow}(\omega),\,\tilde{c}_{\mathbf{k}\uparrow}^{\dagger}(-\omega),\,\tilde{c}_{-\mathbf{k}\downarrow}^{\dagger}(-\omega)\big)^{T},\\
\tilde{\mathbf{b}}_{x,\mathbf{k}}(\omega) & =\big(\tilde{b}_{x,\mathbf{k}\uparrow}(\omega),\,\tilde{b}_{x,\mathbf{k}\downarrow}(\omega),\,\tilde{b}_{x,\mathbf{k}\uparrow}^{\dagger}(-\omega),\,\,\tilde{b}_{x,\mathbf{k}\downarrow}^{\dagger}(-\omega)\big)^{T}.\nonumber
\end{align}
Then the Fourier images of the dynamical equations (\ref{eq:Heisenberg})
become ($\omega^{+}:=\omega+i\epsilon$)
\begin{align}
-i\omega^{+}\tilde{\mathbf{d}}(\omega) & =\hat{\mathbf{d}}(0)-i\mathbf{H}_{\text{w}}\cdot\tilde{\mathbf{d}}(\omega)+i\sum_{\mathbf{k}}[\mathbf{T}_{\mathbf{k}}]_{4N\times4}\cdot\tilde{\mathbf{c}}_{\mathbf{k}}(\omega)+i\sum_{x=1,N}\sum_{\mathbf{k}}[\mathbf{Y}_{x,\mathbf{k}}]_{4N\times4}\cdot\tilde{\mathbf{b}}_{x,\mathbf{k}}(\omega),\nonumber \\
-i\omega^{+}\tilde{\mathbf{c}}_{\mathbf{k}}(\omega) & =\hat{\mathbf{c}}_{\mathbf{k}}(0)-i\mathbf{H}_{\mathrm{sc}}(\mathbf{k})\cdot\tilde{\mathbf{c}}_{\mathbf{k}}(\omega)+i[\mathbf{T}_{\mathbf{k}}^{\dagger}]_{4\times4N}\cdot\tilde{\mathbf{d}}(\omega),\\
-i\omega^{+}\tilde{\mathbf{b}}_{x,\mathbf{k}}(\omega) & =\hat{\mathbf{b}}_{x,\mathbf{k}}(0)-i\mathbf{H}_{\text{e-}x}(\mathbf{k})\cdot\tilde{\mathbf{b}}_{x,\mathbf{k}}(\omega)+i[\mathbf{Y}_{x,\mathbf{k}}^{\dagger}]_{4\times4N}\cdot\tilde{\mathbf{d}}(\omega).\nonumber
\end{align}
 Here $\mathbf{T}_{\mathbf{k}}$ and $\mathbf{Y}_{x,\mathbf{k}}$
are the the $4N\times4$ tunneling matrices indicating the coupling
with the \emph{s}-wave SC and lead-$x$ respectively, and they are
defined by
\begin{equation}
\mathbf{T}_{\mathbf{k}}:=\big[\mathsf{T}_{1,\mathbf{k}},\mathsf{T}_{2,\mathbf{k}},\dots,\mathsf{T}_{N,\mathbf{k}}\big]^{T},\qquad\mathbf{Y}_{1,\mathbf{k}}:=\big[\mathsf{Y}_{1,\mathbf{k}},\mathbf{0},\dots,\mathbf{0}\big]^{T},\qquad\mathbf{Y}_{N,\mathbf{k}}:=\big[\mathbf{0},\dots,\mathbf{0},\mathsf{Y}_{N,\mathbf{k}}\big]^{T},\label{eq:T-matrix}
\end{equation}
 with $4\times4$ blocks $\mathsf{T}_{n,\mathbf{k}}:=\mathrm{diag}\big\{\mathtt{J}_{n,\mathbf{k}},\mathtt{J}_{n,\mathbf{k}},-\mathtt{J}_{n,-\mathbf{k}}^{*},-\mathtt{J}_{n,-\mathbf{k}}^{*}\big\}$
and $\mathsf{Y}_{x,\mathbf{k}}:=\mathrm{diag}\big\{\mathtt{g}_{x,\mathbf{k}},\mathtt{g}_{x,\mathbf{k}},-\mathtt{g}_{x,\mathbf{k}}^{*},-\mathtt{g}_{x,\mathbf{k}}^{*}\big\}$.

The dynamics for the \emph{s}-wave SC and the two electron leads can
be formally obtained with the help of Green functions
\begin{alignat}{2}
\tilde{\mathbf{c}}_{\mathbf{k}}(\omega)= & \mathsf{G}_{\mathrm{sc}}(\omega,\mathbf{k})\cdot\big[\hat{\mathbf{c}}_{\mathbf{k}}(0)+i\mathbf{T}_{\mathbf{k}}^{\dagger}\cdot\tilde{\mathbf{d}}(\omega)\big], & \mathsf{G}_{\mathrm{sc}}(\omega,\mathbf{k}):= & i[\omega^{+}-\mathbf{H}_{\mathbf{k}}^{\mathrm{sc}}]_{4\times4}^{-1},\nonumber \\
\tilde{\mathbf{b}}_{x,\mathbf{k}}(\omega)= & \mathsf{G}_{\text{e-}x}(\omega,\mathbf{k})\cdot\big[\hat{\mathbf{b}}_{x,\mathbf{k}}(0)+i\mathbf{Y}_{x,\mathbf{k}}^{\dagger}\cdot\tilde{\mathbf{d}}(\omega)\big],\qquad & \mathsf{G}_{\text{e-}x}(\omega,\mathbf{k}):= & i[\omega^{+}-\mathbf{H}_{\mathbf{k}}^{\text{e-}x}]_{4\times4}^{-1}.\label{eq:Green-sc}
\end{alignat}
The Green function of lead-$x$ is $\mathsf{G}_{\text{e-}x}(\omega)=i\,\mathrm{diag}\big\{(\omega^{+}-\varepsilon_{x,\mathbf{k}})^{-1},(\omega^{+}-\varepsilon_{x,\mathbf{k}})^{-1},(\omega^{+}+\varepsilon_{x,\mathbf{k}})^{-1},(\omega^{+}+\varepsilon_{x,\mathbf{k}})^{-1}\big\}$,
while $\mathsf{G}_{\mathrm{sc}}(\omega)$ for the \emph{s}-wave SC
requires calculating the inverse of the Hamiltonian matrix $\mathbf{H}_{\mathbf{k}}^{\mathrm{sc}}$.
Taking $\tilde{\mathbf{c}}_{\mathbf{k}}(\omega)$, $\tilde{\mathbf{b}}_{x,\mathbf{k}}(\omega)$
back into the equation of $\tilde{\mathbf{d}}(\omega)$, the dynamics
of the nanowire becomes
\begin{equation}
-i\omega^{+}\tilde{\mathbf{d}}(\omega)=\hat{\mathbf{d}}(0)-i\mathbf{H}_{\mathrm{w}}\cdot\tilde{\mathbf{d}}(\omega)-[\tilde{\mathbf{D}}_{\mathrm{sc}}(\omega)+\tilde{\mathbf{D}}_{\mathrm{e}}(\omega)]\cdot\tilde{\mathbf{d}}(\omega)+i\tilde{\boldsymbol{\xi}}_{\mathrm{sc}}(\omega)+i\tilde{\boldsymbol{\xi}}_{\mathrm{e}}(\omega).\label{eq:QLE-w}
\end{equation}
where $\tilde{\boldsymbol{\xi}}_{\mathrm{sc}}(\omega)$, $\tilde{\boldsymbol{\xi}}_{\mathrm{e}}(\omega)$
are the random forces, $\tilde{\mathbf{D}}_{\mathrm{sc}}(\omega)$,
$\tilde{\mathbf{D}}_{\mathrm{e}}(\omega)$ are the dissipation kernels,
and they are given by
\begin{alignat}{2}
\tilde{\boldsymbol{\xi}}_{\mathrm{sc}}(\omega) & :=\sum_{\mathbf{k}}\mathbf{T}_{\mathbf{k}}\cdot\mathsf{G}_{\mathrm{sc}}(\omega,\mathbf{k})\cdot\hat{\mathbf{c}}_{\mathbf{k}}(0),\qquad & \tilde{\boldsymbol{\xi}}_{\mathrm{e}}(\omega) & :=\tilde{\boldsymbol{\xi}}_{1}+\tilde{\boldsymbol{\xi}}_{N}:=\sum_{x=1,N}\sum_{\mathbf{k}}\mathbf{Y}_{x,\mathbf{k}}\cdot\mathsf{G}_{\text{e-}x}(\omega,\mathbf{k})\cdot\hat{\mathbf{b}}_{x,\mathbf{k}}(0),\nonumber \\
\tilde{\mathbf{D}}_{\mathrm{sc}}(\omega) & :=\sum_{\mathbf{k}}\mathbf{T}_{\mathbf{k}}\cdot\mathsf{G}_{\mathrm{sc}}(\omega,\mathbf{k})\cdot\mathbf{T}_{\mathbf{k}}^{\dagger}, & \tilde{\mathbf{D}}_{\mathrm{e}}(\omega) & :=\tilde{\mathbf{D}}_{\text{e-}1}+\tilde{\mathbf{D}}_{\text{e-}N}:=\sum_{x=1,N}\sum_{\mathbf{k}}\mathbf{Y}_{x,\mathbf{k}}\cdot\mathsf{G}_{\text{e-}x}(\omega,\mathbf{k})\cdot\mathbf{Y}_{x,\mathbf{k}}^{\dagger}.\label{eq:randomForce}
\end{alignat}
In the time domain, Eq.\,(\ref{eq:QLE-w}) gives the quantum Langevin
equation in the main text
\begin{equation}
\partial_{t}\hat{\mathbf{d}}=\hat{\mathbf{d}}(0)\delta(t)-i\mathbf{H}_{\mathrm{w}}\cdot\hat{\mathbf{d}}(t)-\int_{0}^{t}\mathrm{d}\tau\,\mathbf{D}(t-\tau)\cdot\hat{\mathbf{d}}(\tau)+i\hat{\boldsymbol{\xi}}_{\mathrm{sc}}(t)+i\hat{\boldsymbol{\xi}}_{\mathrm{e}}(t),
\end{equation}
where $\mathbf{D}(t):=\mathbf{D}_{\mathrm{sc}}(t)+\mathbf{D}_{\mathrm{e}}(t)$,
and $\hat{\mathbf{d}}(0)\delta(t)$ brings in the initial condition
(omitted in the main text). Then the nanowire dynamics can be obtained
from the Langevin equation (\ref{eq:QLE-w}), i.e.,
\begin{align}
\tilde{\mathbf{d}}(\omega) & =\mathbf{G}(\omega)\cdot\big[\hat{\mathbf{d}}(0)+i\tilde{\boldsymbol{\xi}}_{\mathrm{sc}}(\omega)+i\tilde{\boldsymbol{\xi}}_{\mathrm{e}}(\omega)\big],\nonumber \\
\mathbf{G}(\omega) & :=i[\omega^{+}-\mathbf{H}_{\mathrm{w}}+i\tilde{\mathbf{D}}_{\mathrm{sc}}(\omega)+i\tilde{\mathbf{D}}_{\mathrm{e}}(\omega)]^{-1},\label{eq:G-w}
\end{align}
 where $[\mathbf{G}(\omega)]_{4N\times4N}$ is the Green function
for the nanowire.

\section{The dissipation kernel and effective interaction}

The dissipation kernels in Eq.\,(\ref{eq:randomForce}) provide both
dissipation effect and effective interaction to the nanowire system.
For the two electron leads, with the help of the Green functions (\ref{eq:Green-sc})
and tunneling matrices (\ref{eq:T-matrix}), the dissipation kernel
$\tilde{\mathbf{D}}_{\mathrm{e}}(\omega)=[\tilde{\mathbf{D}}_{\text{e-}1}+\tilde{\mathbf{D}}_{\text{e-}N}](\omega)$
is given by
\begin{align}
\tilde{\mathbf{D}}_{\text{e-}1}(\omega) & =\sum_{\mathbf{k}}\mathbf{Y}_{1,\mathbf{k}}\cdot\mathsf{G}_{\text{e-}1}(\omega,\mathbf{k})\cdot\mathbf{Y}_{1,\mathbf{k}}^{\dagger}=\mathrm{diag}\big\{\tilde{\mathsf{D}}_{\text{e-}1}(\omega),\mathbf{0},\dots,\,\mathbf{0}\big\},\qquad\tilde{\mathbf{D}}_{\text{e-}N}=\mathrm{diag}\big\{\mathbf{0},\dots,\mathbf{0},\tilde{\mathsf{D}}_{\text{e-}N}\big\},\nonumber \\
\tilde{\mathsf{D}}_{\text{e-}x}(\omega) & :=\sum_{\mathbf{k}}\mathsf{Y}_{x,\mathbf{k}}\cdot\mathsf{G}_{\text{e-}x}(\omega,\mathbf{k})\cdot\mathsf{Y}_{x,\mathbf{k}}^{\dagger}=\sum_{\mathbf{k}}\big|\mathtt{g}_{x,\mathbf{k}}\big|^{2}\,\mathrm{diag}\big\{\frac{i}{\omega^{+}-\varepsilon_{x,\mathbf{k}}},\frac{i}{\omega^{+}-\varepsilon_{x,\mathbf{k}}},\frac{i}{\omega^{+}+\varepsilon_{x,\mathbf{k}}},\frac{i}{\omega^{+}+\varepsilon_{x,\mathbf{k}}}\big\}.\label{eq:De-x}
\end{align}
The above summation over the electron modes $\mathbf{k}$ can be turned
into an integral by introducing $\Upsilon_{x}(\omega)$ as the spectral
density of the coupling strength between the nanowire and lead-$x$,
i.e.,
\begin{align}
\Upsilon_{x}(\omega) & :=2\pi\sum_{\mathbf{k}}|\mathtt{g}_{x,\mathbf{k}}|^{2}\delta(\omega-\varepsilon_{x,\mathbf{k}})\nonumber \\
\sum_{\mathbf{k}}\frac{i|\mathtt{g}_{x,\mathbf{k}}|^{2}}{\omega^{+}\pm\varepsilon_{x,\mathbf{k}}} & \longrightarrow i\int_{-\infty}^{+\infty}\frac{\mathrm{d}\varepsilon}{2\pi}\,\frac{\Upsilon_{x}(\varepsilon)}{\omega+i\epsilon\pm\varepsilon}=i\mathcal{P}\int\frac{\mathrm{d}\varepsilon}{2\pi}\,\frac{\Upsilon_{x}(\varepsilon)}{\omega\pm\varepsilon}+\frac{1}{2}\Upsilon_{x}(\pm\omega).\label{eq:Sum-k}
\end{align}

The first term of principle integral provides an energy correction
to the system Hamiltonian $\mathbf{H}_{\mathrm{w}}$ in the Green
function (\ref{eq:G-w}), and the second term provides the dissipation
effect. In transport measurements, the coupling spectral density is
usually approximated as a constant $\Upsilon_{x}(\omega)\equiv\Upsilon_{x}$,
which is also known as the wide band limit. Thus, the above energy
correction gives zero, only left $\tilde{\mathsf{D}}_{\text{e-}x}(\omega)\simeq\frac{1}{2}\Upsilon_{x}\mathrm{diag}\big\{1,1,1,1\big\}:=\frac{1}{2}\mathsf{\Gamma}_{x}$.
As a result, the dissipation kernel of the two electron leads only
provides the dissipation effect, i.e., $\tilde{\mathbf{D}}_{\mathrm{e}}(\omega)\simeq\frac{1}{2}\tilde{\boldsymbol{\Gamma}}_{\mathrm{e}}(\omega):=\frac{1}{2}(\boldsymbol{\Gamma}_{1}+\boldsymbol{\Gamma}_{N})$,
where $\boldsymbol{\Gamma}_{1}:=\mathrm{diag}\big\{\mathbf{0},\dots,\mathbf{0},\mathsf{\Gamma}_{1}\big\}$,
$\boldsymbol{\Gamma}_{N}:=\mathrm{diag}\big\{\mathbf{0},\dots,\mathbf{0},\mathsf{\Gamma}_{N}\big\}$
are $4N\times4N$ dissipation matrices.

Similarly, the dissipation kernel from the \emph{s}-wave SC {[}Eq.\,(\ref{eq:randomForce}){]}
is treated in the same way, which gives
\begin{align}
\tilde{\mathbf{D}}_{\mathrm{sc}}(\omega) & =\sum_{\mathbf{k}}\mathbf{T}_{\mathbf{k}}\cdot\mathsf{G}_{\mathrm{sc}}(\omega,\mathbf{k})\cdot\mathbf{T}_{\mathbf{k}}^{\dagger}\simeq\mathrm{diag}\big\{\tilde{\mathsf{D}}_{1}^{\mathrm{s}}(\omega),\,\dots,\,\tilde{\mathsf{D}}_{N}^{\mathrm{s}}(\omega)\big\},\nonumber \\
\tilde{\mathsf{D}}_{n}^{\mathrm{s}}(\omega) & :=\sum_{\mathbf{k}}\mathsf{T}_{n,\mathbf{k}}\cdot\frac{i}{\omega^{+}-\mathbf{H}_{\mathrm{sc}}(\mathbf{k})}\cdot\mathsf{T}_{n,\mathbf{k}}^{\dagger}\equiv\tilde{\mathsf{D}}_{\mathrm{s}}(\omega).\label{eq:Dsc-4N}
\end{align}
Here, the off-diagonal blocks in $[\tilde{\mathbf{D}}_{\mathrm{sc}}(\omega)]_{4N\times4N}$
are omitted for the local tunneling approximation, which means the
tunneling processes from different sites of the nanowire to the \emph{s}-wave
SC do not have any interferences or correlations with each other.
Besides, since the coupling strengths between different sites and
the \emph{s}-wave SC have the same amplitude, $|\mathtt{J}_{m,\mathbf{k}}|=|\mathtt{J}_{n,\mathbf{k}}|:=\mathtt{J}_{\mathbf{k}}$
for $m\neq n$, all the $N$ diagonal blocks in $\tilde{\mathbf{D}}_{\mathrm{sc}}(\omega)$
are equal to each other, and they are given by
\begin{align}
 & \tilde{\mathsf{D}}_{\mathrm{s}}(\omega)=\sum_{\mathbf{k}}\mathsf{T}_{\mathbf{k}}\cdot\frac{i}{\omega^{+}-\mathbf{H}_{\mathrm{sc}}(\mathbf{k})}\cdot\mathsf{T}_{\mathbf{k}}^{\dagger}=\sum_{\mathbf{k}}\frac{i|\mathtt{J}_{\mathbf{k}}|^{2}}{(\omega^{+})^{2}-(\epsilon_{\mathbf{k}}^{\mathrm{sc}})^{2}-\Delta_{\mathrm{s}}^{2}}\left[\begin{array}{cccc}
\omega^{+}+\epsilon_{\mathbf{k}}^{\mathrm{sc}} &  &  & -\Delta_{\mathrm{s}}\\
 & \omega^{+}+\epsilon_{\mathbf{k}}^{\mathrm{sc}} & \Delta_{\mathrm{s}}\\
 & \Delta_{\mathrm{s}} & \omega^{+}-\epsilon_{\mathbf{k}}^{\mathrm{sc}}\\
-\Delta_{\mathrm{s}} &  &  & \omega^{+}-\epsilon_{\mathbf{k}}^{\mathrm{sc}}
\end{array}\right],\\
 & \frac{1}{(\omega^{+})^{2}-E^{2}}=\frac{1}{2E}\Big[\frac{1}{\omega^{+}-E}-\frac{1}{\omega^{+}+E}\Big]=\frac{1}{2E}\Big[\mathcal{P}\frac{1}{\omega-E}-i\pi\delta(\omega-E)-\mathcal{P}\frac{1}{\omega+E}+i\pi\delta(\omega+E)\Big],\nonumber
\end{align}
where $\mathsf{T}_{\mathbf{k}}:=|\mathtt{J}_{\mathbf{k}}|\,\mathrm{diag}\big\{1,1,-1,-1\big\}$.
Similarly like Eq.\,(\ref{eq:Sum-k}), the summation over $\mathbf{k}$
can be turned into an integral by introducing $\Upsilon_{\mathrm{s}}(\omega)$
as the the spectral density of the coupling strength between the nanowire
and the \emph{s}-wave SC, i.e.,
\begin{equation}
\Upsilon_{\mathrm{s}}(\omega):=\pi\sum_{\mathbf{k}}|\mathtt{J}_{\mathbf{k}}|^{2}\delta(\omega-\epsilon_{\mathbf{k}}^{\mathrm{sc}})\sim\pi|\mathtt{J}_{\mathrm{s}}(\omega)|^{2}\rho_{\mathrm{s}}(\omega).
\end{equation}
Here, $\rho_{\mathrm{s}}(\omega)$ is the density of state from the
\emph{s}-wave SC, and approximately $\Upsilon_{\mathrm{s}}(\omega)\simeq\Upsilon_{\mathrm{s}}$
is a constant. Then the dissipation kernel $\tilde{\mathsf{D}}_{\mathrm{s}}(\omega)$
from the \emph{s}-wave SC is obtained as
\begin{align}
\tilde{\mathsf{D}}_{\mathrm{s}}(\omega) & =\frac{1}{2}\tilde{\mathsf{\Gamma}}_{\mathrm{s}}(\omega)+i\tilde{\mathsf{V}}_{\mathrm{s}}(\omega),\nonumber \\
\tilde{\mathsf{V}}_{\mathrm{s}}(\omega) & =-\frac{\Theta(\Delta_{\mathrm{s}}-|\omega|)\Upsilon_{\mathrm{s}}}{\sqrt{\Delta_{\mathrm{s}}^{2}-\omega^{2}}}\left[\begin{array}{cccc}
\omega &  &  & -\Delta_{\mathrm{s}}\\
 & \omega & \Delta_{\mathrm{s}}\\
 & \Delta_{\mathrm{s}} & \omega\\
-\Delta_{\mathrm{s}} &  &  & \omega
\end{array}\right],\qquad\tilde{\mathsf{\Gamma}}_{\mathrm{s}}(\omega)=\frac{2\Theta(|\omega|-\Delta)\Upsilon_{\mathrm{s}}}{\sqrt{\omega^{2}-\Delta^{2}}}\,|\omega|\mathbf{1}_{4\times4},\label{eq:Ds}
\end{align}
 where $\tilde{\mathsf{\Gamma}}_{\mathrm{s}}(\omega)$ indicates the
dissipation effect, while $\tilde{\mathsf{V}}_{\mathrm{s}}(\omega)$
can be regarded as an effective interaction in the Green function
(\ref{eq:G-w}). Correspondingly, the full dissipation kernel (\ref{eq:Dsc-4N})
can be written as $\tilde{\mathbf{D}}_{\mathrm{sc}}(\omega):=\frac{1}{2}\tilde{\boldsymbol{\Gamma}}_{\mathrm{s}}(\omega)+i\tilde{\mathbf{V}}_{\mathrm{s}}(\omega)$,
with $\tilde{\mathbf{V}}_{\mathrm{s}}(\omega):=\mathrm{diag}\{\tilde{\mathsf{V}}_{\mathrm{s}},\dots,\tilde{\mathsf{V}}_{\mathrm{s}}\}$
and $\tilde{\boldsymbol{\Gamma}}_{\mathrm{s}}(\omega):=\mathrm{diag}\{\tilde{\mathsf{\Gamma}}_{\mathrm{s}},\dots,\tilde{\mathsf{\Gamma}}_{\mathrm{s}}\}$.

It is worth noticing that the Heaviside function appears in both $\tilde{\mathsf{\Gamma}}_{\mathrm{s}}(\omega)$
and $\tilde{\mathsf{V}}_{\mathrm{s}}(\omega)$. That means, when the
system energy lies within the SC gap $|\omega|<\Delta_{\mathrm{s}}$,
the \emph{s}-wave SC only provides the effective pairing (the SC proximity)
without any dissipation effect; for the high energy modes outside
the SC gap $|\omega|>\Delta_{\mathrm{s}}$, the \emph{s}-wave SC only
gives the dissipation effect but does not induce the SC proximity.

\section{Steady state current}

Here we consider the electric current flowing from lead-$1$ to the
nanowire, and the differential conductance measurement. Generally,
the electric current can be calculated by the changing rate of the
electron number in lead-$1$, that is,
\begin{equation}
\langle\hat{I}_{1}(t)\rangle=-e\sum_{\mathbf{k}s}\partial_{t}\langle\hat{b}_{1,\mathbf{k}s}^{\dagger}\hat{b}_{1,\mathbf{k}s}\rangle=\frac{ie}{\hbar}\sum_{\mathbf{k}s}\Big[\mathtt{g}_{1,\mathbf{k}}\langle\hat{d}_{1s}^{\dagger}(t)\hat{b}_{1,\mathbf{k}s}(t)\rangle-\mathtt{g}_{1,\mathbf{k}}^{*}\langle\hat{b}_{1,\mathbf{k}s}^{\dagger}(t)\hat{d}_{1s}(t)\rangle\Big]:=\mathtt{I}_{1}(t)+\text{c.c.}\label{eq:I(t)}
\end{equation}
In particular, we focus on the steady state current after a long enough
time relaxation $t\rightarrow\infty$. This can be obtained from the
Fourier image $\tilde{\mathtt{I}}_{1}(\omega)$ based on the \emph{final
value theorem}, which gives $\mathtt{I}_{1}(t\rightarrow+\infty)=\lim_{\omega\rightarrow0}[-i\omega\tilde{\mathtt{I}}_{1}(\omega)]$.
With the help of the tunneling matrix $\mathbf{Y}_{1,\mathbf{k}}$
and a projector operator $\mathbf{P}^{+}$ defined below, $\tilde{\mathtt{I}}_{1}(\omega)$
can be rewritten as the following matrix form
\begin{align}
\tilde{\mathtt{I}}_{1}(\omega)= & \frac{ie}{\hbar}\sum_{\mathbf{k}}\int\frac{\mathrm{d}\nu}{2\pi}\,\big\langle[\tilde{\mathbf{d}}(\nu)]^{\dagger}\cdot\mathbf{P}^{+}\cdot\mathbf{Y}_{1,\mathbf{k}}\cdot\tilde{\mathbf{b}}_{1,\mathbf{k}}(\nu+\omega)\big\rangle\nonumber \\
= & \frac{ie}{h}\int\mathrm{d}\nu\,\Big\langle\big[-i\tilde{\boldsymbol{\xi}}_{\mathrm{sc}}^{\dagger}(\nu)-i\tilde{\boldsymbol{\xi}}_{\mathrm{e}}^{\dagger}(\nu)\big]\cdot\mathbf{G}^{\dagger}(\nu)\cdot\mathbf{P}^{+}\cdot\big[\tilde{\boldsymbol{\xi}}_{1}(\nu+\omega)+i\tilde{\mathbf{D}}_{\text{e-}1}(\nu+\omega)\cdot\tilde{\mathbf{d}}(\nu+\omega)\big]\Big\rangle\nonumber \\
= & \frac{e}{h}\int\mathrm{d}\nu\,\Big\langle\big\{[\tilde{\boldsymbol{\xi}}_{\mathrm{sc}}^{\dagger}+\tilde{\boldsymbol{\xi}}_{1}^{\dagger}+\tilde{\boldsymbol{\xi}}_{N}^{\dagger}]\cdot\mathbf{G}^{\dagger}\big\}_{(\nu)}\cdot\mathbf{P}^{+}\cdot\big\{\tilde{\boldsymbol{\xi}}_{1}-\tilde{\mathbf{D}}_{\text{e-}1}\cdot\mathbf{G}\cdot[\tilde{\boldsymbol{\xi}}_{\mathrm{sc}}+\tilde{\boldsymbol{\xi}}_{1}+\tilde{\boldsymbol{\xi}}_{N}]\big\}_{(\nu+\omega)}\Big\rangle.\label{eq:I(w)}
\end{align}
Here the projector $\mathbf{P}^{+}$ is defined as $\mathbf{P}^{+}:=\mathrm{diag}\{\mathsf{P}^{+},\mathsf{P}^{+},\dots,\mathsf{P}^{+}\}$,
with blocks $\mathsf{P}^{+}:=\mathrm{diag}\{1,1,0,0\}$. The dynamics
of $\tilde{\mathbf{b}}_{1,\mathbf{k}}(\nu+\omega)$, $[\tilde{\mathbf{d}}(\nu)]^{\dagger}$
has been given in Eqs.\,(\ref{eq:Green-sc}, \ref{eq:G-w}), and
the initial states of the nanowire like $\hat{\mathbf{d}}(0)$ do
not appear here since their contributions would decay to zero in the
steady state $t\rightarrow\infty$.

The quantum expectations in Eq.\,(\ref{eq:I(w)}) are calculated
from the random forces $\tilde{\boldsymbol{\xi}}_{\mathrm{sc}}$,
$\tilde{\boldsymbol{\xi}}_{\mathrm{e}}$ based on the initial states
of lead-$1,N$ and the \emph{s}-wave SC, and they can be expressed
by the correlation matrices for these three fermionic baths, $[\mathbb{C}_{1}(\bar{\omega},\omega)]_{mn}:=\big\langle[\tilde{\boldsymbol{\xi}}_{1}^{\dagger}(\bar{\omega})]_{n}[\tilde{\boldsymbol{\xi}}_{1}(\omega)]_{m}\big\rangle$,
$[\mathbb{C}_{\mathrm{s}}(\bar{\omega},\omega)]_{mn}:=\big\langle[\tilde{\boldsymbol{\xi}}_{\mathrm{sc}}^{\dagger}(\bar{\omega})]_{n}[\tilde{\boldsymbol{\xi}}_{\mathrm{sc}}(\omega)]_{m}\big\rangle$,
which further give
\begin{align}
\tilde{\mathtt{I}}_{1}(\omega)=\frac{e}{h}\int\mathrm{d}\nu\, & \mathrm{tr}\big[\mathbb{C}_{1}(\nu,\omega+\nu)\cdot\mathbf{G}^{\dagger}(\nu)\cdot\mathbf{P}^{+}\big]-\sum_{y=1,N}\mathrm{tr}\big[\mathbb{C}_{y}(\nu,\omega+\nu)\cdot\mathbf{G}^{\dagger}(\nu)\cdot\mathbf{P}^{+}\cdot\tilde{\mathbf{D}}_{\text{e-}1}(\nu+\omega)\cdot\mathbf{G}(\nu+\omega)\big]\nonumber \\
- & \mathrm{tr}\big[\mathbb{C}_{\mathrm{s}}(\nu,\omega+\nu)\cdot\mathbf{G}^{\dagger}(\nu)\cdot\mathbf{P}^{+}\cdot\tilde{\mathbf{D}}_{\text{e-}x}(\nu+\omega)\cdot\mathbf{G}(\nu+\omega)\big].\label{eq:I(w)-C}
\end{align}
According to the final value theorem, with the help of the correlation
matrices $\mathbb{C}_{x,\mathrm{s}}(\bar{\omega},\omega)$ calculated
in Sec.\,\ref{sec:Correlation-matrices}, in the steady state $t\rightarrow\infty$,
the electric current is obtained as
\begin{align}
\langle\hat{I}_{1}\rangle_{\infty}=\frac{e}{h}\int\mathrm{d}\nu\, & \mathrm{tr}\big[\mathbf{G}^{\dagger}\boldsymbol{\Gamma}_{1}^{+}\mathbf{G}\boldsymbol{\Gamma}_{N}^{+}\big]_{(\nu)}\,\big(f_{1}(\nu)-f_{N}(\nu)\big)+\mathrm{tr}\big[\mathbf{G}^{\dagger}\boldsymbol{\Gamma}_{1}^{+}\mathbf{G}\boldsymbol{\Gamma}_{N}^{-}\big]_{(\nu)}\,\big(f_{1}(\nu)-\bar{f}_{N}(-\nu)\big)\nonumber \\
+ & \mathrm{tr}\big[\mathbf{G}^{\dagger}\boldsymbol{\Gamma}_{1}^{+}\mathbf{G}\boldsymbol{\Gamma}_{1}^{-}\big]_{(\nu)}\,\big(f_{1}(\nu)-\bar{f}_{1}(-\nu)\big)+\mathrm{tr}\big[\mathbf{G}^{\dagger}\boldsymbol{\Gamma}_{1}^{+}\mathbf{G}\tilde{\boldsymbol{\Gamma}}_{\mathrm{s}}\big]_{(\nu)}\,f_{1}(\nu)\nonumber \\
- & \mathrm{tr}\big[\mathbf{G}^{\dagger}\boldsymbol{\Gamma}_{1}^{+}\mathbf{G}\tilde{\mathbf{K}}_{\mathrm{s}}^{+}\big]_{(\nu)}\,f_{\mathrm{s}}(\nu)-\mathrm{tr}\big[\mathbf{G}^{\dagger}\boldsymbol{\Gamma}_{1}^{+}\mathbf{G}\tilde{\mathbf{K}}_{\mathrm{s}}^{-}\big]_{(\nu)}\,\bar{f}_{\mathrm{s}}(-\nu).\label{eq:current}
\end{align}
Here the dot symbols for the matrix product `` $\cdot$ '' are omitted
for short. $f_{x}(\nu)$, $f_{\mathrm{s}}(\nu)$ are the Fermi distributions
of lead-$x$ and the \emph{s}-wave SC, and $\bar{f}_{x,\mathrm{s}}(\omega):=1-f_{x,\mathrm{s}}(\omega)$.
The dissipation matrices $\tilde{\boldsymbol{\Gamma}}_{\mathrm{s}}(\nu)$,
$\boldsymbol{\Gamma}_{x}$ have been given in Eqs.\,(\ref{eq:De-x},
\ref{eq:Sum-k}, \ref{eq:Ds}). $\boldsymbol{\Gamma}_{x}^{\pm}$ and
$\tilde{\mathbf{K}}_{\mathrm{s}}^{\pm}(\nu)$ are dissipation matrices
of lead-$x$ and the \emph{s}-wave SC, which are given in Eqs.\,(\ref{eq:Cx-Gamma-x},
\ref{eq:Cs-Ks}). The following relation is needed when deriving the
above result,
\begin{equation}
\tilde{\mathbf{G}}+\tilde{\mathbf{G}}^{\dagger}=\tilde{\mathbf{G}}^{\dagger}\cdot[i(\omega-\mathbf{H}-i\tilde{\mathbf{D}})-i(\omega-\mathbf{H}+i\tilde{\mathbf{D}})]\cdot\tilde{\mathbf{G}}=\tilde{\mathbf{G}}^{\dagger}\cdot(\boldsymbol{\Gamma}_{1}+\boldsymbol{\Gamma}_{N}+\tilde{\boldsymbol{\Gamma}}_{\mathrm{s}})\cdot\tilde{\mathbf{G}}.
\end{equation}

For a transport measurement, we set the chemical potentials as $\mu_{N}=\mu_{\mathrm{sc}}=0$,
$\mu_{1}=eV$, with $V$ as the voltage bias. At the zero temperature,
the chemical potential of lead-$1$ is $f_{1}(\nu)=\Theta(eV-\nu)$,
thus the above electric current (\ref{eq:current}) gives the differential
conductance $\sigma\equiv dI_{1}/dV$ as
\begin{equation}
\sigma=\frac{e^{2}}{h}\Big\{\mathrm{tr}\big[\mathbf{G}^{\dagger}\boldsymbol{\Gamma}_{1}^{+}\mathbf{G}\boldsymbol{\Gamma}_{N}\big]_{(eV)}+\mathrm{tr}\big[\mathbf{G}^{\dagger}\boldsymbol{\Gamma}_{1}^{+}\mathbf{G}\tilde{\boldsymbol{\Gamma}}_{\mathrm{s}}\big]_{(eV)}+\mathrm{tr}\big[\mathbf{G}^{\dagger}\boldsymbol{\Gamma}_{1}^{+}\mathbf{G}\boldsymbol{\Gamma}_{1}^{-}\big]_{(eV)}+\mathrm{tr}\big[\mathbf{G}^{\dagger}\boldsymbol{\Gamma}_{1}^{+}\mathbf{G}\boldsymbol{\Gamma}_{1}^{-}\big]_{(-eV)}\Big\}.
\end{equation}
The first two terms indicate the contributions from the electron exchange
from lead-1 to lead-$N$ and the \emph{s}-wave SC. The last two terms
come from the Andreev reflection between lead-1 and the nanowire.
In the above derivations, except the local tunneling approximation
and the constant coupling spectrum, no other approximations are made.
Thus, in principle this result also applies for the situations when
the magnetic field or coupling strength is strong.

\section{Correlation matrices of the baths \label{sec:Correlation-matrices}}

Here we calculate the correlation matrices in the above current (\ref{eq:I(w)-C}).
From the random force (\ref{eq:randomForce}), Green function (\ref{eq:Green-sc}),
and the tunneling matrix (\ref{eq:T-matrix}), the $4N\times4N$ correlation
matrix $[\mathbb{C}_{x}(\bar{\omega},\omega)]_{mn}:=\big\langle[\tilde{\boldsymbol{\xi}}_{x}^{\dagger}(\bar{\omega})]_{n}[\tilde{\boldsymbol{\xi}}_{x}(\omega)]_{m}\big\rangle$
for lead-$x$ is
\begin{align}
\mathbb{C}_{1}(\bar{\omega},\omega) & =\mathrm{diag}\big\{\mathsf{C}_{1}(\bar{\omega},\omega),\,\mathbf{0},\,\dots,\,\mathbf{0}\big\},\qquad\mathbb{C}_{N}(\bar{\omega},\omega)=\mathrm{diag}\big\{\mathbf{0},\,\dots,\,\mathbf{0},\,\mathsf{C}_{N}(\bar{\omega},\omega)\big\},\nonumber \\{}
[\mathsf{C}_{x}(\bar{\omega},\omega)]_{4\times4} & =\sum_{\mathbf{k}}\mathrm{diag}\big\{\frac{|\mathtt{g}_{x,\mathbf{k}}|^{2}f_{x}(\varepsilon_{\mathbf{k}})}{(\bar{\omega}^{-}-\varepsilon_{\mathbf{k}})(\omega^{+}-\varepsilon_{\mathbf{k}})},\,\frac{|\mathtt{g}_{x,\mathbf{k}}|^{2}f_{x}(\varepsilon_{\mathbf{k}})}{(\bar{\omega}^{-}-\varepsilon_{\mathbf{k}})(\omega^{+}-\varepsilon_{\mathbf{k}})},\,\frac{|\mathtt{g}_{x,\mathbf{k}}|^{2}\bar{f}_{x}(\varepsilon_{\mathbf{k}})}{(\bar{\omega}^{-}+\varepsilon_{\mathbf{k}})(\omega^{+}+\varepsilon_{\mathbf{k}})},\,\frac{|\mathtt{g}_{x,\mathbf{k}}|^{2}\bar{f}_{x}(\varepsilon_{\mathbf{k}})}{(\bar{\omega}^{-}+\varepsilon_{\mathbf{k}})(\omega^{+}+\varepsilon_{\mathbf{k}})}\big\}\nonumber \\
 & =\mathrm{diag}\big\{\frac{i\Upsilon_{x}(\omega)f_{x}(\omega)}{(\omega-\bar{\omega})+2i\epsilon},\,\frac{i\Upsilon_{x}(\omega)f_{x}(\omega)}{(\omega-\bar{\omega})+2i\epsilon},\,\frac{i\Upsilon_{x}(-\bar{\omega})\bar{f}_{x}(-\bar{\omega})}{(\omega-\bar{\omega})+2i\epsilon},\,\frac{i\Upsilon_{x}(-\bar{\omega})\bar{f}_{x}(-\bar{\omega})}{(\omega-\bar{\omega})+2i\epsilon}\big\}.
\end{align}
Here $f_{x}(\omega)$ is the Fermi distribution from the initial equilibrium
state of lead-$x$, i.e.,
\begin{equation}
f_{x}(\omega)=\langle\hat{b}_{x,\mathbf{k}}^{\dagger}(0)\hat{b}_{x,\mathbf{k}}(0)\rangle=\frac{1}{e^{(\omega-\mu_{x})/T_{x}}+1}\stackrel{T_{x}\rightarrow0}{\longrightarrow}\Theta(\mu_{x}-\omega),\qquad\bar{f}_{x}(\omega):=1-f_{x}(\omega).
\end{equation}
The above summations of $\mathbf{k}$ are turned into integrals by
using the coupling spectral density $\Upsilon_{x}(\omega)\equiv\Upsilon_{x}$,
i.e.,
\begin{align}
\sum_{\mathbf{k}}\frac{|\mathtt{g}_{x,\mathbf{k}}|^{2}f_{x}(\varepsilon_{\mathbf{k}})}{(\bar{\omega}^{-}-\varepsilon_{\mathbf{k}})(\omega^{+}-\varepsilon_{\mathbf{k}})} & \rightarrow\int\frac{\mathrm{d}\nu}{2\pi}\,\frac{\Upsilon_{x}(\nu)f_{x}(\nu)}{(\nu-\bar{\omega}+i\epsilon)(\nu-\omega-i\epsilon)}=\frac{i\Upsilon_{x}(\omega)f_{x}(\omega)}{(\omega-\bar{\omega})+2i\epsilon},\nonumber \\
\sum_{\mathbf{k}}\frac{|g_{x}(\mathbf{k})|^{2}\bar{f}_{x}(\varepsilon_{\mathbf{k}})}{(\bar{\omega}^{-}+\varepsilon_{\mathbf{k}})(\omega^{+}+\varepsilon_{\mathbf{k}})} & \rightarrow\int\frac{\mathrm{d}\nu}{2\pi}\,\frac{\Upsilon_{x}(\nu)\bar{f}_{x}(\nu)}{(\nu+\bar{\omega}-i\epsilon)(\nu+\omega+i\epsilon)}=\frac{i\Upsilon_{x}(-\bar{\omega})\bar{f}_{x}(-\bar{\omega})}{(\omega-\bar{\omega})+2i\epsilon}.
\end{align}

Therefore, when calculating the steady state current from the final
value theorem, the correlation matrix in Eq.\,(\ref{eq:I(w)-C})
gives
\begin{align}
\lim_{\omega\rightarrow0}\big[(-i\omega)\mathsf{C}_{x}(\nu,\omega+\nu)\big] & =\Upsilon_{x}\,\mathrm{diag}\big\{ f_{x}(\nu),\,f_{x}(\nu),\,\bar{f}_{x}(-\nu),\,\bar{f}_{x}(-\nu)\big\}:=f_{x}(\nu)\,\mathsf{\Gamma}_{x}^{+}+\bar{f}_{x}(-\nu)\,\mathsf{\Gamma}_{x}^{-},\nonumber \\
\lim_{\omega\rightarrow0}\big[(-i\omega)\mathbb{C}_{x}(\nu,\omega+\nu)\big] & =f_{x}(\nu)\,\boldsymbol{\Gamma}_{x}^{+}+\bar{f}_{x}(-\nu)\,\boldsymbol{\Gamma}_{x}^{-},\label{eq:Cx-Gamma-x}
\end{align}
 where $\mathsf{\Gamma}_{x}^{+}:=\Upsilon_{x}\,\mathrm{diag}\{1,1,0,0\}$
and $\mathsf{\Gamma}_{x}^{-}:=\Upsilon_{x}\,\mathrm{diag}\{0,0,1,1\}$
are upper and lower parts of the dissipation matrix $\mathsf{\Gamma}_{x}\equiv\mathsf{\Gamma}_{x}^{+}+\mathsf{\Gamma}_{x}^{-}$
respectively, and correspondingly $\boldsymbol{\Gamma}_{1}^{\pm}:=\mathrm{diag}\{\mathsf{\Gamma}_{N}^{\pm},\mathbf{0},\dots,\mathbf{0}\}$,
$\boldsymbol{\Gamma}_{N}^{\pm}:=\mathrm{diag}\{\mathbf{0},\dots,\mathbf{0},\mathsf{\Gamma}_{N}^{\pm}\}$.

On the other hand, to calculated the correlation matrix for the \emph{s}-wave
SC, we need to use the Bogoliubov eigen modes, which determine the
initial Fermi distribution of the \emph{s}-wave SC. The \emph{s}-wave
SC Hamiltonian is diagonalized as $\hat{H}_{\mathrm{sc}}=\frac{1}{2}\sum_{\mathbf{k}}\hat{\mathbf{c}}_{\mathbf{k}}^{\dagger}\cdot\mathbf{H}_{\mathbf{k}}^{\mathrm{sc}}\cdot\hat{\mathbf{c}}_{\mathbf{k}}\equiv\frac{1}{2}\sum_{\mathbf{k}}\hat{\boldsymbol{\eta}}_{\mathbf{k}}^{\dagger}\cdot\mathbf{E}_{\mathbf{k}}^{\mathrm{sc}}\cdot\hat{\boldsymbol{\eta}}_{\mathbf{k}}$,
with $\hat{\boldsymbol{\eta}}_{\mathbf{k}}:=\mathsf{U}_{\mathbf{k}}\cdot\hat{\mathbf{c}}_{\mathbf{k}}$
and
\begin{equation}
\mathbf{E}_{\mathbf{k}}^{\mathrm{sc}}:=\left[\begin{array}{cccc}
E_{\mathbf{k}}^{\mathrm{sc}}\\
 & E_{\mathbf{k}}^{\mathrm{sc}}\\
 &  & -E_{\mathbf{k}}^{\mathrm{sc}}\\
 &  &  & -E_{\mathbf{k}}^{\mathrm{sc}}
\end{array}\right]=\mathsf{U}_{\mathbf{k}}\cdot\mathbf{H}_{\mathbf{k}}^{\mathrm{sc}}\cdot\mathsf{U}_{\mathbf{k}}^{\dagger},\qquad\mathsf{U}_{\mathbf{k}}=\left[\begin{array}{cccc}
\cos\theta_{\mathbf{k}} &  &  & \sin\theta_{\mathbf{k}}\\
 & \cos\theta_{\mathbf{k}} & -\sin\theta_{\mathbf{k}}\\
 & \sin\theta_{\mathbf{k}} & \cos\theta_{\mathbf{k}}\\
-\sin\theta_{\mathbf{k}} &  &  & \cos\theta_{\mathbf{k}}
\end{array}\right],\label{eq:Utran}
\end{equation}
where $E_{\mathbf{k}}^{\mathrm{sc}}\equiv[(\epsilon_{\mathbf{k}}^{\mathrm{sc}})^{2}+\Delta_{\mathrm{s}}^{2}]^{1/2}$,
and $\tan2\theta_{\mathbf{k}}\equiv\Delta_{\mathrm{s}}/\epsilon_{\mathbf{k}}^{\mathrm{sc}}$
{[}see also Eq.\,(\ref{eq:bogoliubov transform}){]}.

From the random force (\ref{eq:randomForce}), the correlation matrix
of the \emph{s}-wave SC gives $[\mathbb{C}_{\mathrm{s}}(\bar{\omega},\omega)]_{mn}:=\big\langle[\tilde{\boldsymbol{\xi}}_{\mathrm{sc}}^{\dagger}(\bar{\omega})]_{n}[\tilde{\boldsymbol{\xi}}_{\mathrm{sc}}(\omega)]_{m}\big\rangle\simeq\mathrm{diag}\big\{\mathsf{C}_{\mathrm{s}}(\bar{\omega},\omega),\dots,\mathsf{C}_{\mathrm{s}}(\bar{\omega},\omega)\big\}$,
which is block-diagonal. Similarly as the treatment to the dissipation
kernel (\ref{eq:Dsc-4N}), the off-diagonal blocks are omitted for
the local tunneling approximation. And the diagonal blocks $\mathsf{C}_{\mathrm{s}}(\bar{\omega},\omega)$
are calculated as
\begin{align}
 & [\mathsf{C}_{\mathrm{s}}(\bar{\omega},\omega)]_{mn}=\sum_{\mathbf{k},ij}\Big\langle\big[\hat{\mathbf{c}}_{\mathbf{k}}^{\dagger}(0)\big]_{i}\big[\mathsf{G}_{\mathrm{sc}}^{\dagger}(\bar{\omega})\mathsf{T}_{\mathbf{k}}^{\dagger}\big]_{in}\cdot\big[\mathsf{T}_{\mathbf{k}}\mathsf{G}_{\mathrm{sc}}(\omega)\big]_{mj}\big[\hat{\mathbf{c}}_{\mathbf{k}}(0)\big]_{j}\Big\rangle\nonumber \\
= & \sum_{\mathbf{k}}\mathsf{T}_{\mathbf{k}}\cdot\mathsf{G}_{\mathrm{sc}}(\omega)\cdot\big[\mathbf{1}-\langle\hat{\mathbf{c}}_{\mathbf{k}}(0)\hat{\mathbf{c}}_{\mathbf{k}}^{\dagger}(0)\rangle\big]\cdot\mathsf{G}_{\mathrm{sc}}^{\dagger}(\bar{\omega})\cdot\mathsf{T}_{\mathbf{k}}^{\dagger}=\sum_{\mathbf{k}}\mathsf{T}_{\mathbf{k}}\cdot\mathsf{G}_{\mathrm{sc}}(\omega)\cdot\mathsf{U}_{\mathbf{k}}^{\dagger}\cdot\big[\mathbf{1}-\langle\hat{\boldsymbol{\eta}}_{\mathbf{k}}(0)\hat{\boldsymbol{\eta}}_{\mathbf{k}}^{\dagger}(0)\rangle\big]\cdot\mathsf{U}_{\mathbf{k}}\cdot\mathsf{G}_{\mathrm{sc}}^{\dagger}(\bar{\omega})\cdot\mathsf{T}_{\mathbf{k}}^{\dagger}\nonumber \\
= & \sum_{\mathbf{k}}\mathsf{T}_{\mathbf{k}}\cdot\mathsf{U}_{\mathbf{k}}^{\dagger}\cdot\frac{i}{\omega^{+}-\mathbf{E}_{\mathbf{k}}^{\mathrm{sc}}}\cdot\mathrm{diag}\big\{ f_{\mathrm{s}}(E_{\mathbf{k}}^{\mathrm{sc}}),f_{\mathrm{s}}(E_{\mathbf{k}}^{\mathrm{sc}}),\bar{f}_{\mathrm{s}}(E_{\mathbf{k}}^{\mathrm{sc}}),\bar{f}_{\mathrm{s}}(E_{\mathbf{k}}^{\mathrm{sc}})\big\}\cdot\frac{-i}{\bar{\omega}^{-}-\mathbf{E}_{\mathbf{k}}^{\mathrm{sc}}}\cdot\mathsf{U}_{\mathbf{k}}\cdot\mathsf{T}_{\mathbf{k}}^{\dagger}\nonumber \\
= & \sum_{\mathbf{k}}\frac{|\mathtt{J}_{\mathbf{k}}|^{2}f_{\mathrm{s}}(E_{\mathbf{k}}^{\mathrm{sc}})/2}{(\omega^{+}-E_{\mathbf{k}}^{\mathrm{sc}})(\bar{\omega}^{-}-E_{\mathbf{k}}^{\mathrm{sc}})}\left[\begin{array}{cccc}
1+\cos2\theta_{\mathbf{k}} &  &  & -\sin2\theta_{\mathbf{k}}\\
 & 1+\cos2\theta_{\mathbf{k}} & \sin2\theta_{\mathbf{k}}\\
 & \sin2\theta_{\mathbf{k}} & 1-\cos2\theta_{\mathbf{k}}\\
-\sin2\theta_{\mathbf{k}} &  &  & 1-\cos2\theta_{\mathbf{k}}
\end{array}\right]\nonumber \\
 & \quad+\frac{|\mathtt{J}_{\mathbf{k}}|^{2}\bar{f}_{\mathrm{s}}(E_{\mathbf{k}}^{\mathrm{sc}})/2}{(\omega^{+}+E_{\mathbf{k}}^{\mathrm{sc}})(\bar{\omega}^{-}+E_{\mathbf{k}}^{\mathrm{sc}})}\left[\begin{array}{cccc}
1-\cos2\theta_{\mathbf{k}} &  &  & \sin2\theta_{\mathbf{k}}\\
 & 1-\cos2\theta_{\mathbf{k}} & -\sin2\theta_{\mathbf{k}}\\
 & -\sin2\theta_{\mathbf{k}} & 1+\cos2\theta_{\mathbf{k}}\\
\sin2\theta_{\mathbf{k}} &  &  & 1+\cos2\theta_{\mathbf{k}}
\end{array}\right]\label{eq:Cs}
\end{align}
where $f_{\mathrm{s}}(E_{\mathbf{k}}^{\mathrm{sc}})$ is the Fermi
distribution for the Bogoliubov eigen modes in the initial state.
The summation over the fermion modes $\mathbf{k}$ can be turned into
an integral with the help of the coupling spectral density $\Upsilon_{\mathrm{s}}(\omega)\equiv\Upsilon_{\mathrm{s}}$.
Further, when calculating the steady state current from the final
value theorem, the correlation matrix $\mathsf{C}_{\mathrm{s}}(\bar{\omega},\omega)$
gives
\begin{align}
\lim_{\omega\rightarrow0}\big[(-i\omega)\mathsf{C}_{\mathrm{s}}(\nu,\omega+\nu)\big] & =\frac{2\Upsilon_{\mathrm{s}}}{\sqrt{\nu^{2}-\Delta_{\mathrm{s}}^{2}}}\big[\Theta(\nu-\Delta_{\mathrm{s}})f_{\mathrm{s}}(\nu)-\Theta(-\Delta_{\mathrm{s}}-\nu)\bar{f}_{\mathrm{s}}(-\nu)\big]\,\left[\begin{array}{cccc}
\nu &  &  & -\Delta_{\mathrm{s}}\\
 & \nu & \Delta_{\mathrm{s}}\\
 & \Delta_{\mathrm{s}} & \nu\\
-\Delta_{\mathrm{s}} &  &  & \nu
\end{array}\right]\nonumber \\
 & :=f_{\mathrm{s}}(\nu)\,\tilde{\mathsf{K}}_{\mathrm{s}}^{+}(\nu)+\bar{f}_{\mathrm{s}}(-\nu)\,\tilde{\mathsf{K}}_{\mathrm{s}}^{-}(\nu),\nonumber \\
\lim_{\omega\rightarrow0}\big[(-i\omega)\mathbb{C}_{\mathrm{s}}(\nu,\omega+\nu)\big] & :=f_{\mathrm{s}}(\nu)\,\tilde{\mathbf{K}}_{\mathrm{s}}^{+}(\nu)+\bar{f}_{\mathrm{s}}(-\nu)\,\tilde{\mathbf{K}}_{\mathrm{s}}^{-}(\nu),\label{eq:Cs-Ks}
\end{align}
 where $\tilde{\mathbf{K}}_{\mathrm{s}}^{\pm}(\nu):=\mathrm{diag}\big\{\tilde{\mathsf{K}}_{\mathrm{s}}^{\pm}(\nu),\tilde{\mathsf{K}}_{\mathrm{s}}^{\pm}(\nu),\dots,\tilde{\mathsf{K}}_{\mathrm{s}}^{\pm}(\nu)\big\}$.
To obtain this result, the following derivation is adopted {[}$\mathscr{F}(x)$
is an arbitrary function{]}:
\begin{align}
 & \lim_{\bar{\omega}\rightarrow\omega}\Big[-i(\omega-\bar{\omega})\int_{-\infty}^{\infty}\frac{\mathrm{d}\varepsilon}{2\pi}\,\frac{\mathscr{F}(\sqrt{\varepsilon^{2}+\Delta_{\mathrm{s}}^{2}})}{(\omega^{+}\pm\sqrt{\varepsilon^{2}+\Delta_{\mathrm{s}}^{2}})(\bar{\omega}^{-}\pm\sqrt{\varepsilon^{2}+\Delta_{\mathrm{s}}^{2}})}\Big]\nonumber \\
= & \lim_{\bar{\omega}\rightarrow\omega}\int_{-\infty}^{\infty}\frac{\mathrm{d}\varepsilon}{2\pi}\,i(\bar{\omega}^{-}-\omega^{+})\Big[\frac{1}{\omega^{+}\pm\sqrt{\varepsilon^{2}+\Delta_{\mathrm{s}}^{2}}}-\frac{1}{\bar{\omega}^{-}\pm\sqrt{\varepsilon^{2}+\Delta_{\mathrm{s}}^{2}}}\Big]\frac{\mathscr{F}(\sqrt{\varepsilon^{2}+\Delta_{\mathrm{s}}^{2}})}{\bar{\omega}^{-}-\omega^{+}}\nonumber \\
= & \int_{-\infty}^{\infty}\mathrm{d}\varepsilon\,\mathscr{F}(\sqrt{\varepsilon^{2}+\Delta_{\mathrm{s}}^{2}})\,\delta(\omega\pm\sqrt{\varepsilon^{2}+\Delta_{\mathrm{s}}^{2}})=2\int_{0}^{\infty}dE\,\frac{E\mathscr{F}(E)\,\delta(\omega\pm E)}{\sqrt{E^{2}-\Delta_{\mathrm{s}}^{2}}}=\mp\frac{\Theta(\mp\omega-\Delta_{\mathrm{s}})}{\sqrt{\omega^{2}-\Delta_{\mathrm{s}}^{2}}}\cdot2\omega\mathscr{F}(\mp\omega).
\end{align}

\begin{widetext}
\end{widetext}
\end{widetext}

\end{document}